\documentclass[4pt,a4paper,twoside,twocolumn,american,prd,nofootinbib,superscriptaddress]{revtex4-1}
\usepackage{graphics}
\usepackage[pdftex]{graphicx}
\usepackage{amsmath, amssymb}
\usepackage{color}
\usepackage[usenames,dvipsnames]{xcolor}

\usepackage{setspace}

\usepackage{latexsym}
\usepackage{amssymb}
\usepackage{graphicx,color}
\usepackage{epstopdf}
\usepackage{makeidx} 

\usepackage{lmodern}

\usepackage[T1]{fontenc}
\usepackage[utf8]{inputenc}
\usepackage{color}
\usepackage{babel}
\usepackage{booktabs}
\usepackage{units}
\usepackage{amsmath}
\usepackage{amssymb}
\usepackage{esint}
\usepackage{subfigure}

\pdfpageheight\paperheight
\pdfpagewidth\paperwidth

\usepackage[unicode=true,pdfusetitle,
 bookmarks=true,bookmarksnumbered=false,bookmarksopen=false,
 breaklinks=false,pdfborder={0 0 0},backref=false,colorlinks=true]
 {hyperref}
\hypersetup{
 citecolor=blue,filecolor=blue,linkcolor=blue,urlcolor=blue}

\def\be{\begin{equation}}
\def\ee{\end{equation}}
\def\bea{\begin{eqnarray}}
\def\eea{\end{eqnarray}}
\def\ba{\begin{array}}
\def\ea{\end{array}}
\newcommand{\lsim}{\,\raise 0.4ex\hbox{$<$}\kern -0.8em\lower 0.62ex\hbox{$\sim$}\,}
\newcommand{\gsim}{\,\raise 0.4ex\hbox{$>$}\kern -0.7em\lower 0.62ex\hbox{$\sim$}\,}

\def\de{\mathrm{DE}}

\definecolor{MyBlue}{rgb}{0,0,1} 

\def\be{\begin{equation}}
\def\ee{\end{equation}}

\def\de F{\delta{F}}

\def\H0{H_{0}}

\def\de{\delta}

\definecolor{orange}{rgb}{1,0.5,0}

\everymath{\displaystyle}
\usepackage[]{units}

\usepackage{natbib}

\begin{document}

\title{Asymptotically Safe Starobinsky Inflation}

\author{Edmund J. Copeland}
\email{ed.copeland@nottingham.ac.uk}
\affiliation{School of Physics \& Astronomy, University of Nottingham, Nottingham, NG7 2RD, United Kingdom}
\author{Christoph Rahmede}
\email{christoph.rahmede@kit.edu}
\affiliation{Karlsruhe Institute of Technology, Institute for Theoretical Physics, 76128 Karlsruhe, Germany}
\author{Ippocratis D. Saltas}
\email{ippocratis.saltas@nottingham.ac.uk}
\affiliation{School of Physics \& Astronomy, University of Nottingham, Nottingham, NG7 2RD, United Kingdom}

\newcommand{\addressSussex}{}



\begin{abstract}
We revisit Starobinsky inflation in a quantum gravitational context, by means of the exact Renormalisation Group (RG). We calculate the non-perturbative beta functions for Newton's `constant' 
$G$ and the dimensionless $R^2$ coupling, and show that there exists an attractive UV fixed point where the latter one vanishes but not the former one,
and we provide the corresponding beta functions. The smallness of the $R^2$ coupling, 
required for agreement with inflationary observables, is naturally ensured by its vanishing at the UV fixed point, ensuring the smallness of the primordial fluctuations, 
as well as providing a theoretical motivation for the initial conditions needed for successful inflation in this context.  
We discuss the corresponding RG dynamics, showing both how inflationary and classical observations define the renormalisation 
conditions for the couplings, and also how the UV regime is connected with lower energies along the RG flow. 
Finally, we discuss the consistency of our results when higher order curvature corrections are included, and show that they are robust to the inclusion of $R^3$ corrections. 
\end{abstract}
\maketitle



\section{Introduction}

The Inflationary paradigm has led to a very successful framework within which we can explain the evolution of our Universe from a Hot Big Bang, in particular it has provided us with a mechanism to generate the primordial fluctuations observed in the Cosmic Microwave Background (CMB). There exist a plethora of models in the literature as can be seen through many of the excellent reviews on the subject \cite{Linde:1984ir, Brandenberger:1984cz, Lyth:1998xn,Lyth:2009zz, Mazumdar:2010sa}. Arguably the first model proposed, certainly the first that did not involve an evolving scalar field was due to Starobinsky \cite{Star-R^2}, and it remains perfectly consistent with the most recent {\it Planck} data \cite{Ade:2015lrj}. It is a particular example of a modified theory of gravity in which the  action can be written as a general function $f(R)$ of the Ricci scalar $R$
\be
S = \int d^{4}x \sqrt{-g} f(R)   \label{action:f(R)}
\ee
where in this case we have 
\be
S = \int d^{4}x \sqrt{-g} \left( \frac{1}{16\pi G}R + \frac{1}{b}R^2 \right), \label{action:R^2}
\ee
with the dimensionless coupling $b$ usually expressed as $b \equiv 6 M^2/m_{p}^2$, with $M$ a constant of mass dimension one, where 
$m_{p} \equiv G^{-1/2}$ is the Planck mass, $G$ is Newton's constant which will become scale dependent and $g$ is the determinant of the metric. 

One of the key features of the Starobinsky action is that inflation is driven purely by the gravitational sector, i.~e. without explicit introduction of new fields apart from the metric. 
Originally, the model was motivated by the one-loop corrections to the Einstein--Hilbert action resulting from vacuum 
quantum fluctuations in the matter sector at sufficiently high energies \cite{Star-R^2,Vilenkin:1985md,Birrell:1982ix}, 
which can be taken into account effectively by adding an $R^2$-term to the gravitational action.  
Recently, possibly motivated by its excellent fit to the CMB data (see for example \cite{Ade:2013uln}) there has been increased interest in the model in the context of supergravity
\cite{Dalianis:2014aya,Ferrara:2013rsa, Kallosh:2013hoa, Kallosh:2013lkr, Ferrara:2013kca,Ellis:2013xoa, Ellis:2013nxa, Farakos:2013cqa,Alexandre:2013nqa,Alexandre:2014lla,Pallis:2013yda,Ellis:2014rja,Basilakos:2015yoa} (see Ref. \cite{Ferrara:2013wka} for a more complete discussion on the subject), as well as in other quantum gravity contexts \cite{Briscese:2012ys, Kiritsis:2013gia, Briscese:2013lna,Bamba:2014mua,Bamba:2014jia,Amoros:2014tha}.

Starobinsky inflation is realised at a sufficiently high curvature regime in the early universe, where the $R^2$ term 
dominates the action, resulting in an unstable 
inflationary period with (quasi-) exponential expansion. As the curvature decreases with time, the Einstein--Hilbert term eventually 
comes to dominate at some lower curvature scale, and inflation ends with a graceful exit 
\cite{Linde:2005ht,Linde:1981mu,Kofman:1997yn,Star-R^2}. 

The main goal of this work is to study the Starobinsky model beyond the semiclassical level by considering quantum fluctuations 
in the gravity sector. In particular, we shall calculate the (non-perturbative) Renormalisation Group (RG) flow for the 
two couplings in the action (\ref{action:R^2}), namely Newton's `constant' $G$ and the $R^2$ coupling $b$. 
We will show that for this action, there exists a non--trivial UV fixed point under the RG, where $G$ is asymptotically safe and the $R^2$ coupling $b$ vanishes, i.e. $b \to 0$. 
The vanishing of the coupling $b$ in the UV turns out to be of great importance for a successful inflationary scenario in this context. It is well known that the coefficient of the curvature squared term in the Starobinsky action has to acquire a particularly large value, of the order $\sim 10^9$, to establish agreement with CMB observations. However, from an effective field theoretical point of view, there is no a--priori reason to expect such a large value for the coupling, and inevitably the question arises of how natural such a fine tuning is. This fine tuning could be motivated in the context of some UV completion for gravity, and in this work we will demonstrate that within the scenario of Asymptotic Safety this is indeed the case.

We will further demonstrate that there naturally exist viable inflationary solutions along the RG flow from the UV to IR, in agreement with the {\it Planck} data. In our analysis we will neglect a running cosmological constant in the action, as we are only interested in the original Starobinsky action ansatz. We will consider the inclusion of the cosmological constant along with higher order curvature operators in a work to follow \cite{CRS-in-prog}.

Let us first introduce the idea of Asymptotic Safety (AS) and briefly discuss the previous 
attempts in the literature to implement inflation in this context. AS, first suggested by Weinberg 
\cite{Weinberg:1980gg}, proposes a UV complete theory for gravity by assuming that (metric) gravity is non--perturbatively 
renormalisable through the existence of a non--trivial (interacting) fixed point under the RG. The existence of such a fixed point, 
with a finite
number of attractive directions, which can be successfully connected to the low-energy regime, then yields a predictive theory for gravity. The number of relevant and irrelevant couplings in the UV is found by studying the RG dynamics in the vicinity of the UV fixed point.  It is important to point out that in that regime, the usual power counting arguments do not apply, since the quantum theory is in principle interacting. What is more, one makes in principle no assumptions about the values of the fixed point(s) and the associated eigenvalues in the UV, in particular no assumptions are made about their smallness. This is an immediate result of the non--perturbative nature 
of the AS scenario. In the context of metric theories of gravity, a number of recent results in the literature have provided strong indicators for the existence of 
such a suitable UV fixed point 
\cite{Reuter:1996cp, Souma:1999at, Lauscher:2001ya,Lauscher:2002sq, Litim:2003vp, Reuter:2001ag, Dou:1997fg, 
Codello:2007bd,Codello:2008vh,Machado:2007ea,Benedetti:2009gn,Benedetti:2009rx,BenedettiCaravelli, Benedetti:2011ct, 
Falls:2013bv,Manrique:2011jc,Eichhorn:2013ug,Demmel:2013myx,Dietz:2012ic,Dietz:2013sba,Rechenberger:2012pm}, for reviews, see 
\cite{Niedermaier:2006wt,Percacci:2007sz,Litim:2008tt,Reuter:2012id}.

As we will also discuss more explicitly in section \ref{sec: betas}, the starting point for calculations in this context is an Exact Renormalisation Group equation for the effective action, which implements the Wilsonian idea of integrating out quantum degrees of freedom. For a RG improved action and in the context of AS, inflation has been shown to work consistently for the case of a canonical scalar field coupled minimally to gravity, 
for particular choices of scalar field potentials \cite{Contillo:2011ag, Kaya:2013bga, Cai_RG, Cai:2013caa}. Whereas scalar field inflation models can be made viable by adjusting potentials, 
this is much more of a challenge when inflation is derived from the gravitational sector alone because possible RG trajectories
might never reach the required relative size of couplings. 
Such difficulties have been found in the approaches considered in the literature to date. 
In Refs. \cite{Wein, Reuter:2005kb}, with and without higher-derivative terms in the action,
it was found that viable inflationary solutions with a sufficient number of e-foldings are hard to realise. Furthermore, in Ref. \cite{Hindmarsh:2012rc} it was shown for a RG improved Einstein--Hilbert action that, 
although the inflationary period was able to generate sufficient number of e-foldings, the primordial fluctuations were too 
large to agree with observations. The nature of the latter result can be traced to the position of the UV fixed point 
for the running of Newton's `constant' $G$ and the cosmological constant $\Lambda$ which yields for the dimensionless combination 
\be
\left( G \times \Lambda \right)_{\text{fixed point}} \sim \mathcal{O}(10^{-3}). \label{GLambda}
\ee
However this combination also sets the scale of the gravitational wave power spectrum which in this case is
clearly too large to agree with CMB observations where one would require the product to be more like $\mathcal{O}(10^{-10})$. A similar result was also found in Ref. \cite{Cai:2011kd} at the 1--loop level. 

In this work, we will show an important new result, namely that the original Starobinsky inflationary scenario can work naturally in the context of AS 
through the existence of an UV fixed point where Newton's coupling $G$ reaches a finite value (asymptotically safe coupling), while the $R^2$ coupling $b$ vanishes, a behavior which is similar to the cases considered in Refs. \cite{Fradkin:1981hx,Fradkin:1981iu,Codello:2006in,Niedermaier:2010zz}. We shall comment on the similarity of our analysis to the previous ones in the literature in Section \ref{sec: betas}.

 In section \ref{sec: betas} we present the non--perturbative beta functions, the associated fixed point structure and RG dynamics, while in section \ref{Beta-solutions} we discuss analytic solutions and the implementation of renormalisation conditions for the gravitational couplings. In section \ref{sec:inflationary-solutions} we proceed to study how the slow--roll inflationary solutions obtained in this context are in agreement with the {\it Planck} observations. In section \ref{sec:higher-order-corrections} we begin the important task of addressing the robustness of the solution to higher order curvature corrections. Our results are summarised and discussed in section \ref{sec:conclusions}, and appendices are included where we present explicit expressions and equations.


\section{The non--perturbative beta functions for the Starobinsky action} \label{sec: betas}

The goal of this section is the presentation of the fundamental ingredient for our subsequent analysis,  the non--perturbative 
beta functions under the RG for the action (\ref{action:R^2}). The basis of the calculation will be an Exact RG equation (ERGE) in a Wilsonian context. After deriving the beta functions, we will show the existence of a new fixed point under the RG, which provides a natural setup for a successful implementation of Starobinsky inflation, on the one hand ensuring the smallness of the Starobinsky coupling, as well as the flatness of the inflationary potential at sufficiently high energies.


For completeness, we shall start by briefly sketching the derivation of the exact RG equation, which will be our fundamental tool in our analysis. It starts with an appropriate path integral given by
\be 
Z[J]=\int Dg_{\mu\nu} \;  e^{  i\left(S[g_{\mu \nu}]+   
\int J^{\mu \nu} g_{\mu \nu} + S_{GF}+S_{gh} + \Delta S_{k}  \right) } \label{PI}
\ee
where $S[g_{\mu \nu}]$ corresponds to the bare gravitational action, $J^{\mu \nu}$ is an appropriate external source with $S_{GF}$, $S_{gh}$ denoting appropriate gauge-fixing and ghost terms respectively. The bare action is further modified by the presence of the scale-dependent infrared regulator $\Delta S_k$. The latter is chosen such as to suppres momenta lower than the infrared cut--off $k$, and the integrating out of degrees of freedom proceeds in a Wilsonian fashion, i.e.  shell by shell in momenta \cite{Gies:2006wv, Reuter:2012id, Pawlowski:2005xe, Litim:2008tt}.

The coarse--grained, effective action $\Gamma_k[g_{\mu \nu}]$ associated with the generating functional (\ref{PI}) can be formally derived through a Legendre transformation, and {\it it can be shown to satisfy an Exact Renormalisation Group equation (ERGE)} \cite{Wett_RG-Eq,Morris:1994ie},
\be
k\partial_{k} \Gamma_k[g_{\mu \nu}] = \frac{1}{2}{\rm Tr }
\left[\left(\Gamma_{k}^{(2)} + R_{k} \right)^{-1}  \partial_t R_{k}\right], \label{WetterichEq}
\ee
with $\Gamma_{k}^{(2)}$ denoting the inverse, full propagator, and $\partial_{k} \equiv \partial/\partial k$, where 
the infrared (IR) cut-off scale $k$ sets the coarse graining or RG scale. In particular, momentum modes below $k$ are suppressed, 
while those above $k$ are not, and therefore integrated out. The regulator $R_{k}$ ensures IR regularisation 
as well as finiteness of the trace under very generic requirements 
\cite{Gies:2006wv, Reuter:2012id, Pawlowski:2005xe, Litim:2008tt}.

Evaluation of the ERGE yields the flow equation for the effective action, which describes the RG dynamics of $\Gamma_{k}[g]$ 
as a function of the scale $k$.

A method for solving the ERGE equation, which is also well--suited for our cosmological problem, is the method of truncation, which starts by assuming a particular form for the effective action in  (\ref{WetterichEq}).  In this context, the RG flow based on the ERGE (\ref{WetterichEq}) for an effective action assuming the general $f(R)$ form in (\ref{action:f(R)}) has been calculated in Refs. 
\cite{Codello:2007bd,Machado:2007ea, Codello:2008vh}, 
for different cut--off schemes
on a Euclidean spherical background, 
and we refer the reader to 
these references for its explicit form and details of the calculation. 

It is instructive at this point to briefly sketch the calculation of the RG flow we will be using. Firstly, to calculate the inverse propagator appearing in (\ref{WetterichEq}) the background field method is employed, splitting the metric field into a background and a perturbation piece as $g_{\mu \nu} = \bar{g}_{\mu \nu}  + \delta g_{\mu \nu}$, with the background assumed to be that of a Euclidian sphere. The metric perturbation is further decomposed into its irreducible scalar, vector and tensor pieces respectively under $SO(5)$, allowing an exact calculation of the propagator around $S_4$. The gauge is chosen to be a Landau type gauge, which in turn introduces appropriate ghost terms in the action. The form of the regulator function $R_k$ is chosen to be the so--called ''optimised" (Litim's) cut--off \cite{Litim_Opt-CO1}, which amounts to a step function regularisation of the trace. The trace over momenta on the right hand side of (\ref{WetterichEq}) is evaluated asymptotically using a heat kernel 
expansion. Technical details about the actual calculation are provided in the appendix. 
The result reads schematically,
\begin{equation}\label{Wett_RG-Eq}
384 \pi^2 \left( \partial_{\mu}\tilde{f} + 4\tilde{f} -  2\tilde{R}\tilde{f}_{\tilde{R}} \right) = 
\frac{d\Gamma_k}{d \mu}[\tilde{f}, \tilde{f}_{\tilde{R}}, \tilde{f}_{\tilde{RR}}, \partial_{\mu}\tilde{f}], 
\end{equation}
with $\mu$ being the ``RG-time" $\mu \equiv \ln k/k_0$ for some reference scale $k_0$ (which will eventually become $m_p$), 
$\tilde{R} \equiv R/k^2$ is the dimensionless Ricci scalar, and we have defined 
$\tilde{f} \equiv f(\tilde{R},\mu)/k^4$, $\tilde{f}_{\tilde{R}}\equiv \partial \tilde{f}(\tilde{R},\mu) /\partial{\tilde{R}}$,
$\tilde{f}_{\tilde{R}\tilde{R}}\equiv \partial^2 \tilde{f}(\tilde{R},\mu) /\partial{\tilde{R}^2}$. 
The right hand side of the flow equation is a non--linear equation in $\tilde{R}$, the couplings and their first 
derivatives with respect to the cut--off scale $k$.

In view of (\ref{Wett_RG-Eq}), and for an effective action associated with a Wick-rotated Starobinsky 
action (\ref{action:R^2}), the two beta functions are
\begin{equation}
k \frac{d}{d k} \widetilde{G}(k) = \beta_{\widetilde{G}}(\widetilde{G},b), \; \; \; k \frac{d}{d k} b(k) = \beta_{b}(\widetilde{G},b), \label{betas_full}
\end{equation}
where we have introduced the dimensionless Newton's coupling $\widetilde{G}(k)\equiv k^2 G(k)$. The beta functions are given explicitly in appendix \ref{App:Explicit-eqs}. 
It is important to notice that, although their derivation assumes a Euclidean signature, their validity is expected to carry over to Lorentzian signatures as well. 
Evidence supporting this expectation has been found in Ref. \cite{Manrique:2011jc}. We further notice that, while at the classical level, the action (\ref{action:f(R)}) can be re-expressed in a dynamically equivalent fashion in the so--called Einstein frame, where the Ricci scalar is minimally coupled to a canonical scalar field \cite{Higgs_Auxiliary_Scalars,BIcknell_Auxiliary_Scalars, Teyssandier&Tourrenc_Auxiliary_Scalars, Whitt_Auxiliary_Scalars}, there is no reason to believe that in principle the two representations are equivalent at the quantum level. Evidence for the non--equivalence of the two frames, and in the context of the exact RG, has been recently found in Ref. \cite{Benedetti:2013nya}.

\begin{figure}
  \includegraphics[scale=0.9]{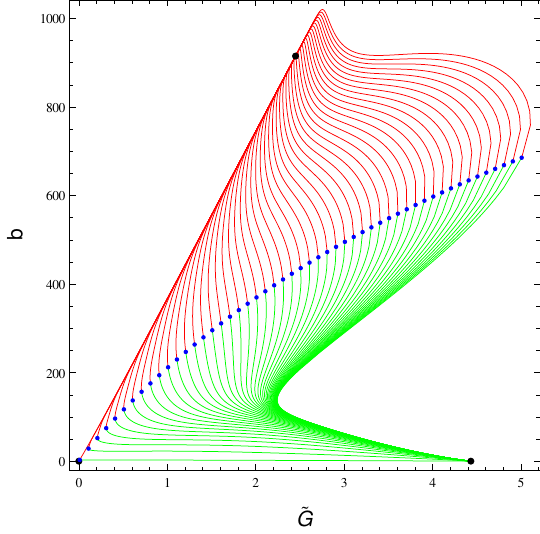}\\
  \caption{\label{plot:Basins} Parametric plot of the RG evolution for the couplings of the Starobinsky action on the $\tilde{G}(k)-b(k)$ plane for different initial conditions. The curves in green correspond to RG trajectories emanating from the fixed point UVFP2, while the ones with red originate from the asymptotically safe fixed point UVFP1. The fixed points are denoted with black dots, while the blue dots correspond to the separatrix between the associated basins of attraction. The fixed point with $\tilde{G}(k)-b(k) = (0,0)$ corresponds to the Gaussian fixed point. Since both UV fixed points are attractive as the cut--off scale is taken to infinity, their basins of attraction are disjoint. Notice that for sufficiently small $\tilde{G}$ ($\ll 1$) at energies well below the Planck mass, the UVFP2 where $b \to 0$ is reached provided $b \ll 1$ as well.
  }
   \end{figure}

At this point we remind the reader that if we were only considering the action (\ref{action:R^2}) with just the  $R^2$ operator, and no linear term, then by itself, it is marginal under standard perturbation theory. The corresponding beta function at leading order in $b$ (derived by sending 
$\widetilde{G} \to 0$ in (\ref{action:R^2})) is then given by
\be
k \frac{d}{d k} b = - \frac{1117}{8640 \pi^2}b^2, \label{beta-R^2}
\ee
and is similar in spirit to the QCD one. The beta function (\ref{beta-R^2}) exhibits one fixed point, $b = 0$, with an associated eigenvalue equal to zero. 
We will show below that the vanishing of the coupling $b$ in the UV persists after the inclusion of the linear curvature term, as described by action (\ref{action:R^2}). Furthermore, we will see that a crucial difference with the beta function (\ref{beta-R^2}) will be the corresponding eigenvalue, which in that case will equal minus two - evidence that we are seeing a pure non-perturbative effect in action here.
The fixed points of the system of beta functions (\ref{betas_full}) can be found by setting the corresponding right hand sides to zero. 
Using the cut--off scheme of Ref. \cite{Codello:2008vh} we find that they exhibit three real-valued fixed points, a ``free", 
Gaussian Fixed Point (GFP), and two ``interacting", UV Fixed Points (UVFP) labeled 1 and 2 respectively, \footnote{To confirm the stability of the UV fixed points under change of cut--off scheme, we checked that using the cut--off scheme of Ref. \cite{Machado:2007ea} instead, the fixed point values show differences only in the fourth decimal place.}
\begin{align}
GFP:  &(\widetilde{G}_{\text fp}, b_{\text fp}) = (0,0), \label{GFP}
\end{align}
\begin{eqnarray}
UVFP1:  \; \; (\widetilde{G}_{\text fp}, b_{\text fp})_{1} &=& (2.451, 914.57), \label{UVFP1}\\
UVFP2: \; \; (\widetilde{G}_{\text fp}, b_{\text fp})_{2} &=& (24\pi/17\approx 4.44, 0). \label{UVFP2}
\label{UVFPs}
\end{eqnarray}
The two non--trivial fixed points correspond to the asymptotic regime of an infinite cut--off scale, $k \to \infty$.
In particular, the second fixed point (\ref{UVFPs}) is rather special, and one of the most important results of this work: 
{\it it describes an asymptotically safe Newton's coupling and a vanishing $R^2$ coupling in the UV}. 
Apart from its interest from a pure RG perspective, this fixed point gives rise to RG trajectories along which  Starobinsky inflation can be viably realised.
On the other hand, at UVFP1, 
both couplings are interacting. We notice that each fixed point gives rise to its own RG dynamics, however, in this work we will be 
mostly interested in UVFP2, which as we will discuss below is crucial for the realisation of Starobinsky inflation in this context. We leave the study of the full theory space to a 
future publication \cite{CRS-in-prog}.
A typical flow from the GFP to UVFP2 according to (\ref{betas_full}) is depicted in Fig. \ref{plot:full-solution}, while the dynamics on the space of couplings in the $\widetilde G-b$ 
plane are shown in Fig. (\ref{plot:Basins}). In particular, the latter plot shows how RG evolution occurs from each fixed point in the deep UV regime towards lower energy scales in the IR. 
Both fixed points are UV attractive (i.e. attractive as the IR cut--off $k$ is taken to infinity), and therefore their basins of attraction are disjoint, a feature which can be explicitly seen 
in Figure (\ref{plot:Basins}). The UVFP2 is an attractor of the RG flow for initial conditions satisfying $\widetilde{G}, b \ll 1$, and as we explain below the physically acceptable 
renormalisation conditions for a viable Starobinsky inflation lie in this range. Therefore, {\it the existence of the fixed point UVFP2 not only ensures the existence of 
RG trajectories along which the primordial fluctuations remain small, but also 
provides a theoretical motivation for the required smallness of the coupling.}

From the linearisation of the flow equations (\ref{betas_full}) around each of the fixed points we obtain the corresponding eigenvalues 
\begin{align}
GFP:    \; \; &(\lambda_{\widetilde{G}}, \, \lambda_b) = (2, 0), \label{GFPEigenvalues}\\
\nonumber \\
UVFP1:  \; \; & (\lambda_{\widetilde{G}}, \, \lambda_b)_{1} = (-39.79, -2.71), 
\label{UVFP1Eigenvalues}  \\
UVFP2:  \; \; & (\lambda_{\widetilde{G}}, \, \lambda_b)_{2}  = (-102/41\approx-2.49, -2). \label{UVFP2Eigenvalues} 
\end{align}

Close to the GFP, perturbation theory applies, and known one-loop corrections to the classical gravitational dynamics have
to be recovered. The GFP eigenvalues (\ref{GFPEigenvalues}) show that the $\widetilde{G}$-direction is UV repulsive, while the $b$- 
one is marginally UV attractive (repulsive) for $b>0$ ($b<0$), a result known from perturbation theory. 

At the same time, the negative eigenvalues (\ref{UVFP1Eigenvalues})--(\ref{UVFP2Eigenvalues}) indicate that both UVFP1 and UVFP2  are fully UV-attractive. As a consequence, there cannot be any RG trajectory connecting UVFP1 and UVFP2 and their respective basins of attraction are disjoint. 

In particular, the non--trivial eigenvalue of the coupling $b$ in (\ref{UVFP2Eigenvalues}) is not what one would infer from a simple dimensional analysis; around the UV fixed point with $b = 0$, the coupling $b$ acquires an anomalous dimension equal to minus two, which is a pure non-perturbative effect. We discuss the origin of this property in appendix \ref{App:Explicit-eqs}. We should notice here that in the previous literature a similar behaviour associated with the curvature-squared sector of gravity has been found and discussed at the 1--loop level, as long as the interaction term of the Weyl tensor contracted with itself was included in the action \cite{Fradkin:1981hx,Fradkin:1981iu,Codello:2006in,Niedermaier:2010zz}. 

{\it The existence of the fixed point  (\ref{UVFPs}) is crucial for the resulting inflationary dynamics
as it provides us with a mechanism for naturally producing small inflationary fluctuations at the perturbative level}. What is more, this type of behaviour is able to overcome the 
previously found problems in the context of AS inflation discussed in the introduction, i.~e. combining a sufficient number of e-folds with the requirement of obtaining the 
correct amplitude for the metric fluctuations.

We are now in a position to discuss the relevant dynamics of the system (\ref{betas_full}). The non--linearity of the beta functions 
allows only for a numerical solution, however close to the fixed point we are most interested in, (i. e. the UVFP2  (\ref{UVFPs})) we obtain an analytic solution by expanding the beta functions around $b=0$, while keeping $\widetilde{G}$ general. 
Doing so, the beta functions organise schematically as 
\begin{align}
& \beta_{\widetilde{G}}(\widetilde{G}, b) \simeq 3 \frac{(17\widetilde{G} - 24\pi )\widetilde{G}}{5\widetilde{G} - 36\pi} + \sum_{n=1}^{\infty} \sum_{m=0}^{2n}C_{(nm)} 
\frac{b^{n} \widetilde{G}^{m-2n+2}}{(5\widetilde{G}-36\pi)^{n+1}}, \label{betag-approx}\\
& \beta_{b}(\widetilde{G},b) \simeq 41\frac{\widetilde{G} \, b}{5\widetilde{G}-36\pi} + \sum_{n=2}^{\infty} \sum_{m=0}^{2n-2}D_{(nm)} \frac{b^{n} \widetilde{G}^{m-2n+3}}
{(5\widetilde{G}-36\pi)^{n}}.\label{betab-approx}
\end{align}
Here $C_{(nm)}$ and $D_{(nm)}$ are constants which can easily be obtained from equations (\ref{betag_coeff})--(\ref{betab_coeff}). 
For the family of RG trajectories we will be interested in ($b \ll 1$), the second and higher order terms in $b$ 
in (\ref{betag-approx})--(\ref{betab-approx}) can be neglected. 
Towards lower cut--off energies, trajectories emerging from UVFP2 will approach the GFP with $\widetilde{G}$ dropping to smaller values, as one can also see from Fig. \ref{plot:full-solution}.
 
As we search for analytic solutions to (\ref{betas_full}) we focus on the 
regime  $b \ll \widetilde{G} \lesssim 1$, as this allows us to further expand  (\ref{betag-approx})--(\ref{betab-approx}) in powers of $\widetilde{G}$. As we shall see, the solutions will allow us to capture very accurately the actual RG flow of the system. To leading order in $\widetilde{G}$ we then obtain,
\begin{align}
&k\frac{ d }{d k}\widetilde{G} \simeq 2\widetilde{G} - \frac{41}{36\pi}\widetilde{G}^2 + \mathcal{O}(\widetilde{G}^3, b),\label{betag-approx2}\\ 
&k\frac{ d }{d k}b \simeq - \frac{41}{36 \pi } \widetilde{G}b + \mathcal{O}(\widetilde{G}^2, b^2). \label{betab-approx2}
\end{align}

The negative sign in (\ref{betab-approx2}) is crucial for the vanishing of the coupling $b$. We will discuss the solutions of the above system in the next section.

Before we close this section, let us remark on two interesting properties of the model (\ref{action:R^2})
in view of the vanishing of the coupling $b$ in the UV. It is well known that models of the $f(R)$ type propagate an extra, 
massive scalar degree of freedom, (often called  a ``scalaron") compared to General Relativity \cite{Sotiriou:2008rp,DeFelice:2010aj}. For the general $f(R)$ action (\ref{action:f(R)})
the effective mass of the scalaron is
\be
m^2_{\text{eff}} \equiv \frac{f_{R}- Rf_{RR}}{3f_{RR}}.
\ee
In terms of the Starobinsky action (\ref{action:R^2}) this corresponds to 
\be
m^2_{\text{eff}} = \frac{m_{p}^2}{96 \pi}b,
\ee
which shows that the value of $m_{\text{eff}}$ compared to $m_p$ at some cut--off scale $k$ is set by $b \equiv b(k)$. 
The point to make here is that, in the limit $k \to \infty$, as the coupling $b \to 0$, 
the scalaron mass vanishes ($m_{\text{eff}} \to 0$), reflecting the absence of any length scales in that limit, and the restoration of scale-invariance. 

What is more, as we explain later in section (\ref{sec:inflationary-solutions}), when the curvature squared term dominates the action (\ref{action:R^2}) the universe expands (quasi-) exponentially. The second fixed point UVFP2 (\ref{UVFP2}) ensures that for a sufficiently high value of the cut--off scale $k$, the coupling $b$ will be sufficiently small for the $R^2$ term to dominate the action and the universe will inflate.

Before we close the section let us comment on the apparent divergence of the action in the limit $b(k) \to 0$ (as $k \to \infty$), which is similar to the case of the vanishing of the 
Yang--Mills coupling in Quantum Chromodynamics. 
In gravity, the $R^2$ term is in that sense analogous being quadratic in curvature and coming with a dimensionless coupling constant $1/b(k)$. 
For a pertubative expansion in $b(k)$ it is convenient to perform a field redefinition in the following way.
Expanding the effective action in powers of the coupling $b(k)$ we would set up a perturbative expansion which close to our UV fixed point UVFP2 would demonstrate that 
the action remains finite in that limit. In particular, one can consider a perturbative expansion of the metric around some background $\bar{g}_{\mu \nu}$, $g_{\mu \nu}=\bar{g}_{\mu \nu}+h_{\mu \nu}$ 
and expand the action in terms of small fluctuations $h_{\mu \nu}$.\footnote{In the derivation of the beta functions, we use a similar background gauge fixing which however does not assume perturbatively
small flucuations $h_{\mu\nu}$.} This would lead to various terms of increasing power of $h_{\mu \nu}$ contracted with each other, the covariant derivative, and background curvature 
tensor terms in a suitable way, corresponding to interactions of increasing order. Now, one can redefine the fluctuating field as $h_{\mu \nu} \to \sqrt{b(k)}h_{\mu \nu}$, and in doing so it becomes 
apparent that the prefactor $1/b(k)$ in the $R^2$ term is then absorbed leaving us with a series of terms 
with increasing positive powers of $b(k)$. 
One has to notice though that after the field redefinition $h_{\mu \nu} \to \sqrt{b(k)}h_{\mu \nu}$, the expansion of the linear curvature term $R$ will contain terms of the form $b(k)/G(k)$, 
and because the dimensionful Newton's coupling $G(k)$ decreases with increasing RG scale $k$ when approaching a UV fixed point, it implies that the perturbative expansion might break down, i.e. that 
we may not always have 
$b(k)/G(k) \ll 1$. However, from the analytic solutions (\ref{g-b-solutions}) below, it follows that in our case the ratio $b(k)/G(k)$ approaches a constant as the RG scale $k$ is sent to infinity, 
since in this limit $G(k) \equiv \widetilde{G}(k)/k^2 \propto 1/k^2$, $b(k) \propto (k_0/k)^{2}$, and therefore the action remains finite.

\section{Solutions and renormalisation conditions} \label{Beta-solutions}
The approximate system of beta functions (\ref{betag-approx2})--(\ref{betab-approx2}) for $\widetilde{G}(k)\equiv k^2 G(k)$ and $b(k)$ can be
solved analytically, after first solving (\ref{betag-approx2}) by separation of variables, and then substituting the solution into (\ref{betab-approx2}), to find
\begin{align}
& \widetilde{G}(k) \simeq  \frac{\widetilde{G}_{0}}{1 + \frac{41\widetilde{G}_0}{72 \pi} (\nicefrac{k}{k_0})^{2}} \left(\nicefrac{k}{k_0}\right)^{2}, \nonumber \\ 
& b(k) \simeq \frac{b_0}{1 + \frac{41\widetilde{G}_0}{72 \pi}  (\nicefrac{k}{k_0})^{2}   }, \label{g-b-solutions}
\end{align}
with $\widetilde{G}_0$ and $b_0$ being constants of integration, and $k_0$ a constant non-vanishing reference scale, which we will choose to be the Planck mass as it is 
measured today, i.~e. $k_0 = m_p$. This is a convenient choice that will allow us to measure everything in units of the Planck mass. 
The values of all physical observables are then defined with respect to the chosen scale.
It is easy to see that in the IR limit, i.e as $k \to 0$, $b\to  b_0$, and $\widetilde{G} \to 0$ respectively.   
This behaviour is in very good agreement with the full numerical solution of equations 
(\ref{betag_coeff})-(\ref{betab_coeff}) (presented in Fig. \ref{plot:full-solution}) in the vicinity of the GFP.

One might worry that the validity of solutions (\ref{g-b-solutions}) breaks down as soon as the two couplings become of 
comparable magnitude, $\widetilde{G} \sim b \ll 1$, i.~e. when the assumption that $b \ll \widetilde{G} \lesssim 1$ is violated. However, even in that case, 
the solutions  (\ref{g-b-solutions}) remain quite accurate. This can be also seen in Fig. \ref{plot:full-solution} where the full, numerical solution, as well as the approximate solutions (\ref{g-b-solutions}) are plotted for appropriate initial conditions. Another way to understand this is by looking at the approximate beta 
functions around the GFP. Notice that according to (\ref{g-b-solutions}), $\widetilde{G}(k)$ changes faster than $b(k)$ due to the 
extra $k^2$ dependence. At the point where $\widetilde{G} \ll b \ll 1$, the beta function $\beta_{b}$, to leading order, will be given 
by (\ref{beta-R^2}), while the one for $\widetilde{G}(k)$ has the same form as (\ref{betag-approx}), but with the $\widetilde{G}^2$-coefficient now being  equal to $\nicefrac{79}{24 \pi}$. The corresponding solutions read as,
\begin{align}
& \widetilde{G}(k) \simeq \frac{\widetilde{G}_{0}}{1 + \frac{79\widetilde{G}_0}{24 \pi} (\nicefrac{k}{k_0})^{2}} \left(\nicefrac{k}{k_0}\right)^{2}, \\
& b(k) \simeq \frac{b_0}{1 + \left( \nicefrac{1117}{8640 \pi^2}\right)b_0 \ln(\nicefrac{k}{k_{0}})}, \label{solution:b-1-loop}
\end{align}
with $b_0 \ll1$. Note the integration constants here $\widetilde{G}_{0}$ and $b_0$ are not the same as in (\ref{g-b-solutions}). 

At the scale where the denominator of (\ref{solution:b-1-loop}) becomes zero, the analytic solution breaks down, and one has to resort to a numerical solution of the full system of beta functions (see also Fig.\ref{plot:full-solution}.) Notice that the pole in (\ref{solution:b-1-loop}) is of similar nature to that of the QCD running coupling at 1--loop. 

The integration constants $\widetilde{G}_0$, $b_0$ in the solutions (\ref{g-b-solutions}) have to be fixed by applying appropriate renormalisation conditions 
at a particular scale $k = k_0$ (recall we will take $k_0 = m_p$). However, an important point is that the measurements available to us for each of these couplings
correspond to different scales (energies). On the one hand, Newton's `constant' $G$ has been measured from micrometer to solar distance scales (low energies), 
while as we shall do in section \ref{sec:inflationary-solutions}, the $R^2$ coupling $b$ should be determined from CMB observations (high energies). 
To overcome this obstacle, our strategy will 
be to match the measured value of each coupling at the corresponding energy scale, $k = k_{\text{measured}}$, through the analytic 
solutions (\ref{g-b-solutions}), and then use (\ref{g-b-solutions}) to extrapolate their values for example to the Planck scale
$k \sim m_p$ as this is the reference scale we have chosen for $k_0$. 

In the classical regime, corresponding to $k\ll k_0=m_p$, we know that $G=\widetilde{G}(k)k^{-2}=1/m_p^2$, hence it follows that we need to choose our initial constant as $\widetilde{G}_0 \simeq 1$.
Using the solution (\ref{g-b-solutions}) to extrapolate up to the Planck scale, it then follows that 
\be\label{g-renorm-condition}
\widetilde{G}(k=m_p)\simeq 0.85\ . 
\ee
Relation (\ref{g-renorm-condition}) will provide our renormalisation condition for $\widetilde{G}$ at the Planck scale. 
Notice that, as the cut--off energy $k$ drops much below the Planck scale
($\nicefrac{k}{m_p} \ll1$), the couplings' evolution enters the classical regime, 
where $\widetilde{G}(k)=  k^{2}G(k)\simeq k^2 m_p^{-2}$
and 
$b\simeq b_{0}$. 
The appropriate value for $b_0$ which will define the renormalisation condition for the coupling $b=b(k)$ will be determined in section \ref{sec:inflationary-solutions}, 
using the recently published {\it Planck} data.

\section{Inflationary solutions} \label{sec:inflationary-solutions}

The solutions for the running couplings (\ref{g-b-solutions}) are functions of the RG scale $k$, and when inserted into the 
action (\ref{action:R^2}) they result in a continuous family of actions parametrized by the value of the RG scale $k$. 
Now, in cosmology, variables and physical quantities generally depend on the space-time coordinates $x^{\mu}$. One is naturally 
led to think that in an expanding universe the process of integrating out degrees of freedom occurs as a function of time
and space, which in turn demands promoting the IR cut--off $k$ to a (monotonic) space-time dependent function, $k = k(t, \bf{x})$. 
Such a procedure has been implemented in a number of ways 
\cite{Wein, Reuter:2005kb, Guberina:2002wt,Shapiro:2004ch,Babic:2004ev,Domazet:2010bk,Bonanno:2001xi,Bonanno:2001hi,Bonanno:2009nj,Bonanno:2010bt,Bonanno:2012jy,Contillo:2011ag,Frolov:2011ys, Cai:2011kd, Hind-Litim_Rahme, Hindmarsh:2012rc,Koch:2010nn,Koch:2014joa}, with one particular example being associating $k^2 \sim H^2$, since the Hubble parameter $H$ naturally provides an IR cut--off. 
Our approach here will be to RG-improve the $k$-dependent action (\ref{action:R^2}) in a covariant way 
\cite{ Bonanno:2012jy,Frolov:2011ys, Hindmarsh:2012rc}. We will do this by relating the cut--off scale $k$ with the Ricci scalar through 
\be
k^2 = \rho R, \label{cut-off-ident1}
\ee
with $\rho > 0$ constant, a choice which is very similar in spirit to the RG improvement of the effective potential in scalar 
field theory  as applied in Ref. \cite{Coleman:1973jx}. The effect of the identification (\ref{cut-off-ident1}) will be to 
implement the running of the gravitational couplings under the RG as non--linear curvature corrections in the original action 
ansatz (\ref{action:R^2}). The dynamics of the resulting action, which is shown below in (\ref{action:R^2-RG-improved}), can then be
treated with classical methods as we shall also see. 

The value of the constant $\rho \equiv k^2/R$ has a particular physical interpretation: its magnitude describes the importance of 
radiative
corrections in the classical equations. In particular, the case of $\rho \ll 1$ implies that higher order curvature corrections 
in the action (\ref{action:R^2}) are important, but at the same time radiative corrections are small; in the opposite case where  $\rho \gg 1$, higher order curvature corrections are negligible, but radiative ones not necessarily \cite{Wein}. 
As also argued in Ref. \cite{Wein}, the optimal case corresponds to $\rho \sim 1$, which is the case when radiative corrections 
start to become important and higher order corrections in the action become negligible. As we shall explain below, our analysis will be almost insensitive to the actual value of $\rho$, unless $\rho$ is tuned to an extreme amount. 

During slow-roll inflation ($\left|\nicefrac{\dot{H}}{H^2}\right|\ll1$) from (\ref{cut-off-ident1}) 
one then has
\be
k^2 = \rho R= 6\rho H^2(2 + \nicefrac{\dot{H}}{H^2}) \simeq 12 H^2. \label{cut-off-ident2}
\ee
One sees that for the choice of $\rho \sim \mathcal{O}(1)$ the Hubble parameter naturally sets the RG cut-off scale, or equivalently the cosmological horizon sets the typical scale of correlations between quantum degrees of freedom. 

Let us point out that one expects that physics should not depend on the particular identification of the cut—off scale, however in the absence of knowledge of the true theory of quantum gravity, 
one has usually to restrict to a particular truncation ansatz, which inevitably introduces some dependence on the cut--off identification. Below, we will explicitly demonstrate that the parameter $\rho$
has an extremely mild effect on our results. As regards the generality of the choice (\ref{cut-off-ident1}), one could think of the cut--off $k$ being proportional to other combinations of curvature 
invariants (appearing with the appropriate powers), instead of just the Ricci scalar $R$. However, for a FRW spacetime, it is easy to see that any curvature combination will reduce to being proportional 
to an appropriate power of the Hubble parameter.
 
Let us now look at the solutions we found for $\widetilde{G}(k)$ and $b(k)$ given in (\ref{g-b-solutions}). We conveniently 
chose $\widetilde{G}_0 = 1$, so that according to the renormalisation condition (\ref{g-renorm-condition}) there are sizable running coupling
corrections at the Planck scale. 
Inserting the cut--off identification (\ref{cut-off-ident1}), as described above, into the solutions (\ref{g-b-solutions})  for $b(k)$ and $\widetilde{G}(k)$, and in turn plugging the latter 
into the action (\ref{action:R^2}), the RG-improved action reads
\begin{widetext}
\be
S_{\text{RG-improved}} = \int d^{4} x \sqrt{-g} \left(  \frac{m_{p}^2}{16 \pi} R + \frac{1}{b_0}R^2 + \frac{1}{b_0} \left(1 - \frac{41 \rho b_0}{1152 \pi^2} \right) \frac{41 \rho}{72 \pi}\frac{R^3}{m_p^2} \right).
\label{action:R^2-RG-improved} 
\ee
\end{widetext}
We notice that in the derivation of (\ref{action:R^2-RG-improved}) we performed a trivial redefinition of the coupling $b_0$ to bring the curvature squared term's coefficient to its usual form. 
The action (\ref{action:R^2-RG-improved}) is {\it the result of the RG-improvement of the original $R^2$ action} 
(\ref{action:R^2}), based on the solutions (\ref{g-b-solutions}), and is similar in spirit to an RG-improvement of a scalar field theory in the local potential approximation.
We see that a key effect of the RG-improvement of the original action is to generate an effective $R^3$ term in addition to the linear and quadratic one, a manifestion of the radiative corrections in this context. What is more, the contribution of the higher order (cubic) correction, compared to the curvature squared one, is controlled by the dimensionless parameter $\rho$ as we anticipated. 
In fact, for the inflationary relevant energy scales, where the energy (or curvature) of the universe decreases compared to $m_p$, $ R \lesssim m_{p}^2$, we require that the cubic term will remain negligible (in an effective field theory sense), and we can use this fact to put a constraint on $\rho$. This translates into requiring that
\be
\left(1 - \frac{41 \rho b_0}{1152 \pi^2} \right) \frac{41 \rho}{72 \pi}  \lesssim 1. \label{Cubic-term-condition}
\ee
The above condition is quadratic in $\rho$ leading to more than one solution. For the curvature regime and value of the initial condition $b_0$ we are interested in, $R/m_{p}^2 \sim 10^{-10}$ and $b_0 \sim 2 \cdot 10^{-9}$ (see below) we find that
\be
\rho \lesssim \frac{1152 \pi ^2}{41b_0}, \label{rho-bound}
\ee
which in turn implies that $\rho$ would have to be tuned to an extremely high value for our subsequent analysis to break down, since $b_0 \ll 1$. At the same time, the  constraint (\ref{rho-bound}) is compatible with the assumption that the IR, Wilsonian cut--off scale $k$ is associated with the universe's horizon during inflation, i.e. $k^2 \sim \mathcal{O}(1) R$.
We further notice that the condition (\ref{Cubic-term-condition}) is closely related to the requirement that in the RG solutions (\ref{g-b-solutions}), 
the second term in the denominator is much less 
than unity, i. e. $(\nicefrac{41\widetilde{G}_0}{72 \pi}) (\nicefrac{k}{m_p})^2 \ll 1$.

It follows, given the above requirements, the RG improved action (\ref{action:R^2-RG-improved}) now simply reduces to the standard Starobinsky form
\be
S_{\text{RG-improved}} \simeq \int d^{4} x \sqrt{-g} \left(
\frac{m_{p}^2}{16 \pi}R + \frac{1}{b_0} R^2\right). \label{action:R^2-RG-improved2}
\ee
The remaining input required for the action (\ref{action:R^2-RG-improved2}) is to determine the initial value of $b_0$, 
and this will be done by considering the inflationary observables at the corresponding high energy scales. In particular, the required value of $b_0$ for successful inflation will define 
its renormalisation condition in the UV, and select the particular family of RG trajectories.  
As we will see, the (very small) observationally required value is in perfect agreement with RG solutions connecting the UV with the IR regime in a viable way (see also Fig.\ref{plot:full-solution}.)

To determine the key inflationary observables for the action (\ref{action:R^2-RG-improved2}) 
and compare them with the {\it Planck} data, we follow a similar analysis to Refs. 
\cite{Star-R^2, Vilenkin:1985md, Mukhanov:1990me, Amend1992PRD45417A, Maeda1988PRD37858M, DeFelice:2010aj}. 

Let us first derive the background equation of motion for the action (\ref{action:R^2-RG-improved2}), or initially for the more general action (\ref{action:f(R)}). Varying with respect to the metric field, and evaluating on the Friedman--Robertson--Walker (FRW) metric,
\be
ds^2 = -dt^2 + a(t)^2{\bf dx}^2 \label{metric:FRW}
\ee
we get for the $00$- component of the Einstein-equations,
\be
\frac{f_{R}}{H^2}(1-\epsilon) + 6f_{RR}\left(4\epsilon + \frac{\dot{\epsilon}}{H} - 2\epsilon^2 \right) - \frac{1}{6}\frac{f}{H^4} = 0, \label{eom: 00}
\ee 
with 
\be 
\epsilon \equiv -\dot{H}/H^2 \label{slow-roll1}
\ee 
the slow-roll parameter, $f_R \equiv df/dR$, $f_{RR} \equiv d^2 f/dR^2$, $R = 6H^2(2-\epsilon)$ and $\dot{} \equiv d/dt$.

For $\epsilon= 0$ and $\dot{\epsilon} = 0$, the solution of equation (\ref{eom: 00}), if it exists, yields an exact de Sitter expansion where $H = H_0 = \text{const.}$, and it is easy to check that an action with $f(R) \propto R^2$ possesses such a solution. In the context of the action (\ref{action:R^2-RG-improved2}), inflation starts off at some sufficiently high curvature scale, where the $R^2$ term dominates, then as the curvature decreases over time, eventually the linear term becomes significant, and inflation ends. 

To describe the inflationary dynamics of (\ref{action:R^2-RG-improved2}) we will be interested in slow-roll 
solutions of (\ref{eom: 00}), (the appropriate $f(R)$ being given by (\ref{action:R^2-RG-improved2})), where $\epsilon^2, \dot{\epsilon}/H \ll \epsilon \ll 1$, $R \simeq 12H^2, 
\dot{R}\simeq -24H^3 \epsilon$ and $R^2 \simeq 144H^4$. Under these assumptions, and neglecting terms of order 
$(\epsilon^2, \nicefrac{\dot{\epsilon}}{H})$, equation (\ref{eom: 00}) can then be integrated to give
\be
H(t) \simeq H_{0} -  \frac{1}{576 \pi}b_{0}m_{p}^2(t - t_0), \label{H(t)}
\ee
with $H_0 \equiv H(t_0) = \text{constant}$, and $t_0$ the time when inflation begins. 
The number of e--foldings $N$ between some scale $H(t)$ and  the end of inflation, is given by
\be
N \equiv \int_{t}^{t_{end}} H(t) dt  = - \int_{H}^{H_{\rm end}} \frac{d \log{H}}{\epsilon} 
\simeq \frac{288\pi}{b_{0}m_{p}^2} ( H^2 - H_{\rm end}^2), \label{N}
\ee
with the slow-roll parameter $\epsilon$ calculated from the solution (\ref{H(t)}) as 
\be
\epsilon\equiv \epsilon(t) = \frac{1}{36} \frac{  \nicefrac{m_{p}^2}{16\pi} }{ \nicefrac{1}{b_0}  } \frac{1}{H(t)^2}, \label{epsilon2}
\ee
and where $H_{\rm end}$ is the value of $H$ when inflation ends which corresponds to when $\epsilon =1$ in (\ref{epsilon2}).
Notice that (\ref{epsilon2}) implies that the universe inflates as long as the $R^2$ term dominates the action (\ref{action:R^2-RG-improved2}).

Small, inhomogeneous fluctuations of the metric field during inflation contribute to the temperature inhomogeneities observed in the CMB. For the case of the action (\ref{action:R^2-RG-improved2}) the cosmologically relevant propagating degrees of freedom around the FRW background (\ref{metric:FRW}) are the two polarisation modes of the transverse-traceless (spin-two) field and a scalar (spin-zero) degree of freedom respectively (see e.g \cite{Mukhanov:1990me} for an explicit analysis.). 

The corresponding power spectra for an $f(R)$ type action have been calculated in Refs. 
\cite{Mukhanov:1990me, PhysRevD.71.063536, PhysRevD.54.1460,DeFelice:2010aj}, and are presented in (\ref{Pscalar-general}) and (\ref{Ptensor-general}) for convenience. From (\ref{Pscalar-general}) and using the slow-roll approximations (\ref{epsilon-approximations}), we obtain for the amplitude of the scalar fluctuations evaluated at horizon crossing $k_{F} = aH$, i.~e. when the particular scale $k_F$ leaves the cosmological horizon \footnote{$k_F$ here stands for the Fourier wavenumber to make the distinction with the RG scale $k$.},
\begin{align}
P_S \simeq \frac{1}{48 \pi^2}\frac{H^2}{f_{R}}\frac{1}{\epsilon^2} \simeq \frac{N^2}{288 \pi^2} b_{0}. \label{Pscalar-approx}
\end{align}
In the last approximate equality we used (\ref{epsilon2}) to relate $\epsilon$ with $N$ at the 
time of horizon crossing of the mode $k_{F}$, by approximating $\epsilon \simeq \nicefrac{1}{2N}$, always in 
the slow-roll regime. Furthermore, we approximated 
$f_{R} = \nicefrac{m_{p}^2}{(16 \pi)} + (\nicefrac{12}{b_0})(2-\epsilon)H^2 \simeq \nicefrac{24}{b_0}H^2$, since as we will see shortly $b_{0} \ll 1$. 

In a similar fashion, from relation (\ref{Ptensor-general}) for the tensor amplitude during slow-roll we find that
\begin{align}
P_{T} \simeq \frac{1}{\pi^2} \frac{H^2}{f_{R}} \simeq \frac{1}{24 \pi^2} b_{0}. \label{Ptensor-approx}
\end{align}
The tensor to scalar ratio is easily found to be
\be
r \equiv \frac{P_{T}}{P_{S}} \simeq 48 \epsilon^2 \simeq \frac{12}{N^2}, \label{r}
\ee
where we again have used (\ref{N}) and (\ref{epsilon2}) in the last approximate equality to relate $\epsilon$ with $N$. Notice that the tensor-to-scalar ratio is suppressed by a factor of $\epsilon^2$.

The spectral indices are defined by equations (\ref{Pscalar-general}) and (\ref{Ptensor-general}) in the appendix, and we will assume 
that they are independent of $k_{F}$,  in other words we assume there is effectively no running of the spectral index. Under the slow-roll conditions described in (\ref{epsilon-approximations}) they are approximated as (\ref{nS})
\be
n_{S} -1 \simeq -4\epsilon, \; \; \; \; n_{T} \simeq 0. \label{ns-approx}
\ee
We need to make connection with observations which we do by using the recent {\it Planck} results combined with the WMAP large scale polarisation likelihood ({\it Planck} + WP) \cite{Ade:2013uln}. At the pivot scale $k_F = 0.002 {\rm Mpc}^{-1}$, for the tensor to scalar ratio they require that $r_{0.002}<0.12$. Similarly, the bound for the spectral index and 
scalar amplitude reads as $n_S = (0.9603 \pm 0.0073)$, and $\ln (10^{10}P_S) = 3.089^{+0.024}_{-0.027}$ respectively. 

The typical value of the minimum number of e-foldings needed to solve the flatness and horizon problem is between $50-60$ \cite{Ade:2013uln}. Here, we shall use the value $N = 55$, which from (\ref{r}) implies that $r\simeq 0.004$, certainly within the {\it Planck} bound. At the same time, from (\ref{r}) we further find that $\epsilon \simeq 0.009$, which from (\ref{ns-approx}) implies that $n_S \simeq 0.964$ also within the observational bound. 

We can now use the {\it Planck} upper bound on $\epsilon$ and the approximate relation (\ref{epsilon2}), to first determine an upper bound
for $b_0$. This bound can then be used to put an upper bound on the 
Hubble parameter and, through (\ref{cut-off-ident2}), on  the cut--off scale $k$ as well.

For $N = 55$ in relation (\ref{Pscalar-approx}), the {\it Planck} observation of $P_{S}$ provides us with the important result, the required value of the 
coupling $b_{0}$,
\be
b_0 
\simeq 2.063\times10^{-9} \label{b-bound-Planck}\ .
\ee 
Using (\ref{b-bound-Planck}) in (\ref{epsilon2}) we get an upper bound for $H$ which can then be used to put an upper 
bound on the cut--off scale $k$ through (\ref{cut-off-ident2}). This way, we find
\be
\frac{H}{m_p} \lesssim 1.126 \times10^{-5} \leftrightarrow \frac{k}{m_p} \lesssim 3.89 \times 10^{-5}, \label{bound-H-k}
\ee
for $N=55$ e-foldings before the end of inflation.

To find the renormalisation condition of $b(k)$ at the Planck scale we use (\ref{b-bound-Planck}) in the analytic solution (\ref{g-b-solutions}) to obtain,
\be
b(k =m_p) = 1.757\times10^{-9}. \label{renorm-cond-b}
\ee
This important relation supplements the similar one for $\widetilde{G}(k)$, given in (\ref{g-renorm-condition}), and together they select 
out the RG trajectory which matches with CMB observations. Notice that from the renormalisation group condition for $b_0$ we find from (\ref{rho-bound}) that the dimensionless parameter $\rho$ has to satify $\rho \lesssim 1.58 \times 10^{11}$.

It has to be stressed that a crucial fact regarding the viability of inflation in this context is the (almost) 
constancy of the coupling $b$ for energies below $m_p$.

This can also be seen from the beta function (\ref{betab-approx2}); in particular, 
for $b \sim 10^{-9}$ and $\widetilde{G}  < 1$, (\ref{betab-approx2}) tells us that 
\be
\left | k\frac{d b}{dk} \right| \simeq 10^{-9}\widetilde{G} \ll 1, \label{b-variation}
\ee
obviously in agreement with the upper and lower limits of the {\it Planck} data. We emphasise that the required 
value for $b$, relation (\ref{b-bound-Planck}), can be achieved for a wide range of cut--off scales, thanks to its vanishing at sufficiently high energies, and its tiny variation for $\nicefrac{k}{m_{p}} \ll 1$ along the RG flow.

What is more, the negligible variation of $b(k)$ for $\nicefrac{k}{m_{p}} \ll 1$ implies that the renormalisation 
condition (\ref{renorm-cond-b}) will also provide a prediction for the value of the coupling $b$ at classical scales. 
Notice that purely from classical considerations, a rather weak low energy bound for $b$ can be found as follows: given that 
the average matter density at a distance $r = 10^{3}r_{\odot}$ from the centre of the sun is  
$\rho_{m} \approx 10^{-24} \nicefrac{g}{cm^3}$, one finds that $R = 8\pi G_{\odot}\rho_{m} \approx 10^{-117}m_{p}^2$, 
which in turn implies that, if $b > 10^{-115}$, the effect of the $R^2$ term is negligible, and General Relativity recovered. Therefore, the requirement of viable inflation in this context, provides with a much stronger bound on the coupling $b$ at classical scales. 

We conclude this section, by noting that the vanishing of the curvature squared coupling, $b$, ensures the smallness of its value in the UV, while the asymptotic safety of Newton's `constant' $G$ ensures the absence of UV infinities on the theory space. What is more, the approximate constancy of the coupling $b$ during inflation further guarantees that the primordial fluctuations remain sufficiently small, allowing for a viable period of inflation driven by the $R^2$ term. 

\section{Stability under higher curvature corrections} \label{sec:higher-order-corrections}
In the previous section we showed how under the RG a fixed point exists for the Starobinsky action, where the coupling of the curvature squared term tends to zero, while Newton's coupling becomes  asymptotically safe. An important question to ask is whether this fixed point is stable under higher order corrections, i.e. whether it still exists {\it if we include higher order operators in the effective action from the onset}? In principle one can consider an expansion of the original effective action in an effective field theoretical fashion as
\be
\Gamma_k = \int \sqrt{g} \left[ \frac{1}{16 \pi G} R + \frac{1}{b} R^2 + \frac{1}{c_{3}} R^3 + \mathcal{O}(R^n) \right], \label{EA-EFT}
\ee
where the couplings are assumed to be dependent on the Wilsonian cut--off $k$, but with no a priori Planck-suppression of higher order terms being assumed. In particular, we shall leave the RG solutions for the couplings to reveal the suppression/dominance of the different operators. 

In this section, we include in the effective action terms up to cubic in the scalar curvature, as shown in (\ref{EA-EFT}), and we demonstrate that a similar fixed point to that already found for the Starobinsky truncation is a solution of the associated non--perturbative RG flow of the effective action (\ref{EA-EFT}). 

\begin{figure}
  \includegraphics[scale=0.5]{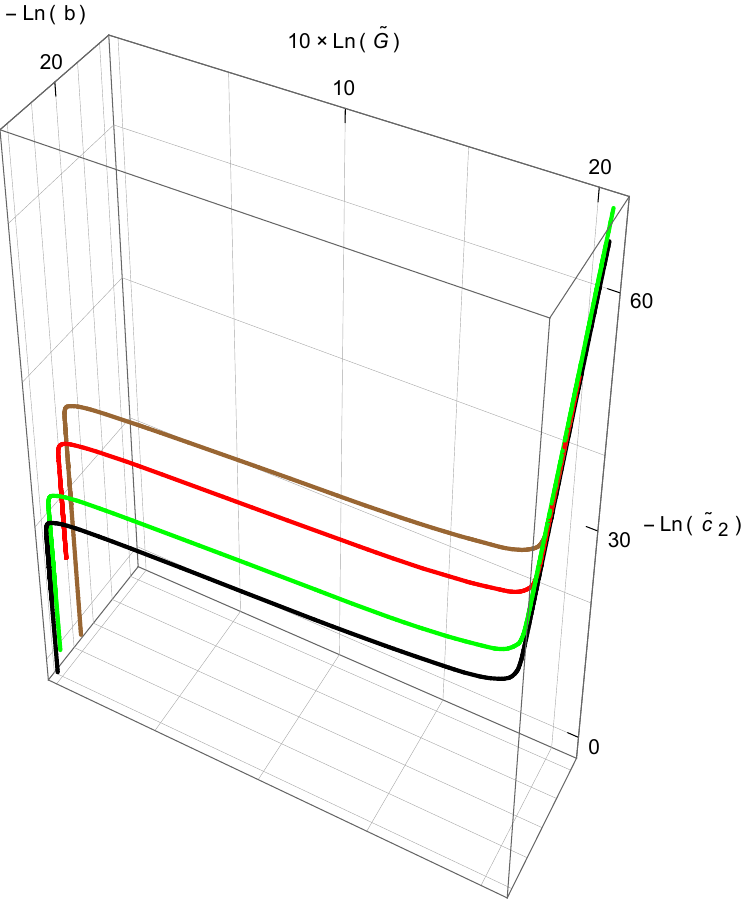}\\
  \caption{\label{plot:3DRcubed} Parametric plot of the non--perturbative RG evolution for the couplings of the truncation emerging from (\ref{EA-EFT}) for the couplings $\widetilde{G}, b$ and $\tilde{c}_{3}$ for different initial conditions. In the UV, Newton's coupling becomes asymptoticaly safe, while the higher order couplings $b$ and $\tilde{c}_3$ flow to zero. Evolution occurs from the top right (corrsponding to $k \to \infty$) and evolves towards the IR as the cut--off scale $k$ decreases. The coupling $\widetilde{G}$ starts off from its non--trivial fixed point value and decreases towards the IR, $b$ remains almost constant, while $\tilde{c}_3$ remains constant for a sufficiently long RG-time before starting to increase in the IR, as one would expect for an IR--irrelevant coupling. Notice that the beta functions are derived from the same RG flow equation as for the Starobinsky action, under the assumption of a Euclidian spherical background spacetime, and the gauge and cut--off choices as explained 
in the appendix. The initial conditions for the different curves brown, red, green and black at $k = m_p$ are $(\widetilde{G}, b, \tilde{c}_{3} ) = \; \;
 (10^{-10},10^{-9} ,10^{-14}), \; \;
(10^{-10},10^{-10},10^{-17}) , \\
(10^{-16},10^{-13},10^{-17}), \,
(10^{-18},10^{-15},10^{-19})$,
 and within the cut--off range $k_{\text{UV}} = 10^{16}m_p$, $k_{\text{IR}} = 10^{-8}m_p$.
 }

   \end{figure}

To derive the full set of non--perturbative beta functions we first plug the action (\ref{EA-EFT}) truncated to cubic order into the ERGE (\ref{WetterichEq}), using the same assumptions of gauge (Landau-type gauge), regulator (optimised regulator) and background ($S_4$) as before, and after the evaluation of the trace we arrive at the full RG flow. Due to the high complexity of the equations we will not give the explicit form of the RG flow or of the associated beta functions here, and we refer to Ref. \cite{Codello:2008vh} for more details. After solving for the beta functions for the three couplings of the action (\ref{EA-EFT}) we find that a similar fixed point to UVFP2, found for the Starobinsky truncation before, extends to this case as well, and reads as
\begin{eqnarray}
(\widetilde{G}_{\text fp}, b_{\text fp}, \tilde{c}_{\text fp\ 3}) = ( 216 \pi/329, 0,0), \label{UVFP-R^3}
\end{eqnarray}
with the (dimensionless)Newton's coupling remaining asymptotically safe, while the ones associated to the curvature square and cubic terms respectively flowing to zero. We have defined the dimensionless cubic coupling as $\tilde{c}_3 \equiv c_3/k^2$. The result of the fixed points (\ref{UVFP-R^3}) is very encouraging. We see that Newtons constant remains close to the value it had previously in (\ref{UVFP2}), the $R^2$ coupling has gone to zero as before and now the potentially troublesome $R^3$ coupling $\tilde c_3$ remains finite in the UV limit. This has the effect of ensuring that the $R^2$ term dominates the effective action in the UV hence guarantees the flatness of the potential even when $R^3$ terms are present.  
Moreover the fixed point (\ref{UVFP-R^3}) is an attractive one, as we now see. The leading order terms in the beta functions sufficiently close to this UV fixed point can be derived from the full RG flow, assuming in a similar manner as before, that $b, \tilde{c}_3 \ll 1$ and $\widetilde G \lesssim 1$. This way we find that
\begin{align}
&k\frac{ d }{d k}\widetilde{G} \simeq 2\widetilde{G} - \frac{329}{108\pi}\widetilde{G}^2 + \mathcal{O}(\widetilde{G}^3, b, \tilde{c}_3),\label{betag-approx3}\\ 
&k\frac{ d }{d k}b \simeq - \frac{443}{23040 \pi^2 } b^2 -  \frac{41}{768 \pi }\frac{b^2}{\widetilde{G}} + \mathcal{O}(\widetilde{G}^2, b^3, \tilde{c}_3),  \label{betab-approx3} \\
& k\frac{ d }{d k}\tilde{c}_3 \simeq  -2\tilde{c}_3 -\frac{2141}{4246732800\pi^3} \frac{\tilde{c}_3^2}{\widetilde{G}}+ \mathcal{O}(\widetilde{G}^2, b^2, \tilde{c}_3^2) \label{betac3-approx3}
\end{align}
It is easy to see that the eigenvalues of the above system of beta functions around the fixed point (\ref{UVFP-R^3}) are $(\lambda_{\widetilde{G}}, \, \lambda_b, \, \lambda_{\tilde{c}_{3}}) = (-2,0,-2)$. The directions associated with the couplings $\widetilde G$ and $\tilde{c}_{3}$ are attractive, while the one associated with $b$ is in this case marginal. 
The RG evolution resulting from the numerical solution of the full system of beta functions for the three couplings from UV to IR is presented in Figure \ref{plot:3DRcubed}. 

The inflationary observables will provide us again with the renormalisation condition for the couplings $b_0$, assuming that the action is dominated by the curvature square term. This provides us with a bound on the initial condition for the coupling $\tilde{c}_{3_0}$,
\be \label{c3-constraint}
\frac{\tilde{c}_{3_0}}{m_{p}^2 b_0} \gg \frac{R_0}{m_{p}^2},
\ee 
with $R_0$ the value of scalar curvature about $55$ e--folds before the end of inflation. The condition (\ref{c3-constraint}) is essentially the requirement that the cubic term in the effective action during inflation is Planck--suppressed compared to the curvature squared one, or in other words that the potential remains sufficiently flat during inflation. 

To conclude this section, we have presented evidence that the crucial fixed point found for the Starobinsky truncation persists even in the presence of a cubic term in the effective action. This is very encouraging and we plan to consider the inclusion of yet higher order terms in the future.

\section{Conclusions and discussion}\label{sec:conclusions}

In this work, we have revisited Starobinsky inflation, by calculating the non-perturbative beta functions for the (Euclidean) vacuum gravitational action (\ref{action:R^2}), in the context of the exact Renormalisation Group (RG) and Asymptotic Safety, the latter assuming a UV completion for the theory under the presence of a UV fixed point under the RG. 
We have presented the full, non-perturbative beta functions for the gravitational couplings (equations (\ref{betag_coeff})-(\ref{betab_coeff})), and derived approximate relations in the UV in 
(\ref{betag-approx})-(\ref{betab-approx}) and (\ref{g-b-solutions}). Although the calculation was performed in Euclidean space, a common choice in these sort of setups, we assume that the 
results extend to Lorentzian signatures as well. Evidence for this fact, and in a similar context to ours, has been found in Ref. \cite{Manrique:2011jc}. Our calculation made use of the 
cut--off of Ref. \cite{Litim_Opt-CO1} which allows for an 
analytic calculation of the flow equation. Although physical observables are expected to be independent of the cut--off function, the particular choice of cut--off function, in combination with 
the truncated theory space, naturally introduces some bias in the results. For our choice of cut--off function (i.e. Litim's ''optimised" cut--off), the latter dependence is expected to be minimised \cite{Litim:2001up,Litim:2000ci}.

The dynamics and predictions of the inflationary regime emerging from the non-trivial UV fixed point found were studied, showing that the vanishing of the curvature squared coupling ensures 
the existence of RG flows from UV to IR along which a successful inflationary period occurs, and this is the case for a wide range of scales along the RG flow. 

\begin{figure} 
\centering
\subfigure[]{
  \includegraphics[scale=0.6]{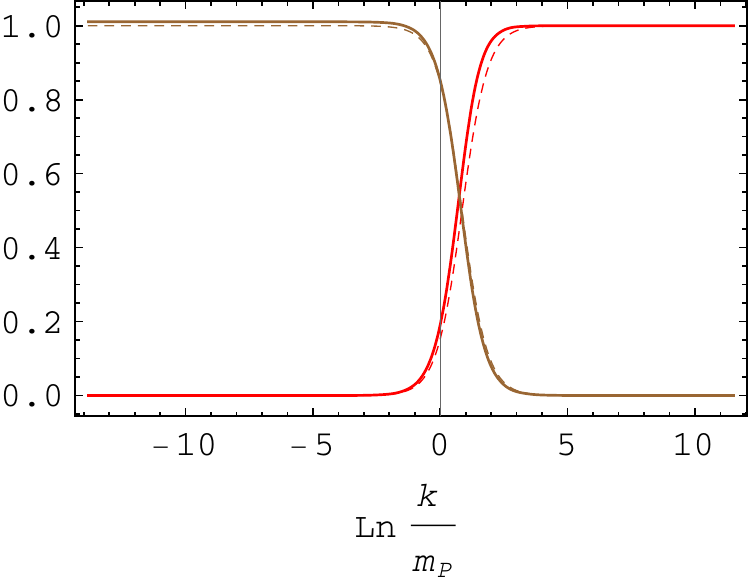}
\label{Subfigure1a}
}

\subfigure[]{
  \includegraphics[scale=0.6]{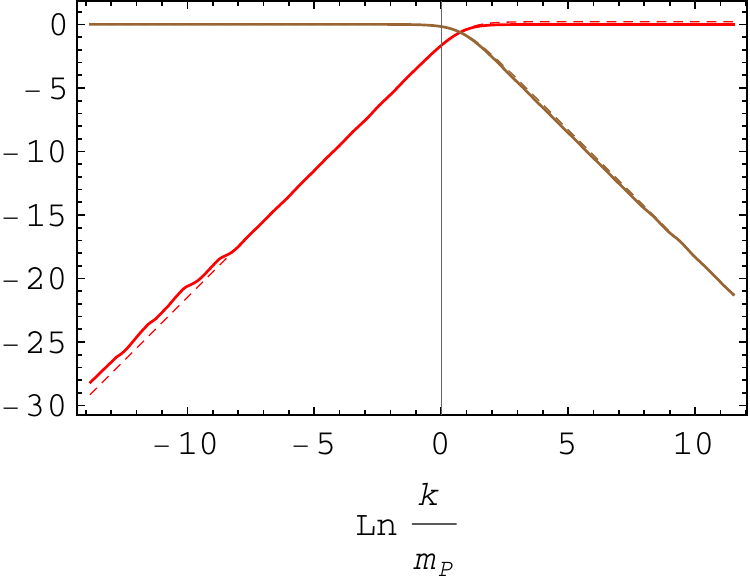}
\label{Subfigure1b}
}
 \caption{\label{plot:full-solution} Numerical solutions of the full system of beta functions (continuous curves), together with the corresponding approximate analytical solutions (\ref{g-b-solutions}) (dashed curves), for the couplings $\widetilde{G}(k)/\widetilde{G}_{\text fp}$ (red) and $b(k)\times 10^{9}/(2.063)$ (brown) in log-linear (upper figure) and log-log space (lower figure), and 
for the initial conditions: $k_{\text{UV}} = 10^{5}m_p$, $k_{\text{IR}} = 10^{-6}m_p$, 
 $\widetilde{G}(k = m_p) = 0.85$, $b(k = m_p) =1.757\times 10^{-9}$. (See also equations (\ref{g-renorm-condition}) and (\ref{renorm-cond-b}) in the text.) The couplings start in the UV close to UVFP2, with $\widetilde{G}_{\text fp} = 24\pi/17$ ($=72\pi/41$ for the analytic solutions (\ref{g-b-solutions})), 
 $b_{\text fp} = 0$, and evolve towards the IR as the cut--off energy $k$ decreases. 
  We expect inflation to occur for energies smaller than the Planck mass, $\nicefrac{k}{m_p} \ll 1$. Notice that in the latter regime, we have approximately $b \simeq \text{const.}$ and $\widetilde{G} \simeq (\text{const.})k^2$. What is more, the lower figure clearly shows the gradients of the curves as they approach their respective fixed-point values, i.e $+2$ for the red, and $-2$ for the brown curve respectively, as one would expect from the analytic solutions (\ref{g-b-solutions}). The vanishing of the coupling $b$ in the UV ensures the smallness of the primordial fluctuations, as well as the flatness of the inflationary potential. 
 }
 \end{figure}

The main results of our analysis can be summarised as follows:

\textbullet $\;$  We showed that under the RG an attractive, non--trivial UV fixed point exists, UVFP2 relation (\ref{UVFP2}), 
where Newton's coupling $G$ is asymptotically safe, and the $R^2$ coupling $b \to 0$. 
We obtained indications that the fixed point is stable under the cut--off scheme adopted, and independent of the gauge
choice due to its origin in the gauge-invariant transverse-traceless part of the flow equation, as discussed in 
section \ref{App:Explicit-eqs} of the appendix.
Furthermore, the fixed point is UV-attractive and is connected with the IR regime along the RG flow, as the cut-off 
energy $k$ decreases, also shown in Fig. \ref{plot:full-solution}. A remarkable property of the new UV fixed point found 
is the non-trivial anomalous dimension ($=-2$) for the dimensionless $R^2$ coupling in its vicinity, which is not expected using 
a naive dimensional analysis. Apart from the non--trivial UVFP2, we also showed that the theory 
space of the action (\ref{action:R^2}) exhibits a trivial, Gaussian fixed point (\ref{GFP}) along with a second, non-trivial UV one (\ref{UVFP1}) disconnected from (\ref{UVFP2}).   

\textbullet $\;$ The fixed point on which the $R^2$ coupling vanishes (UVFP2), $b(k \to \infty) = 0$, ensures the existence of RG trajectories along which the universe enters 
into a de Sitter-like expansion at some sufficiently high cut--off scale, as the $R^2$ term comes to dominate the action 
($1/b(k) \gg 1$). As the cut--off scale $k$ decreases along the RG flow, the curvature also decreases until eventually the 
Einstein-Hilbert term in the action comes to dominate and inflation ends. 

\textbullet $\;$ The UV fixed point UVFP2 further ensures the existence of RG trajectories connecting the UV with the IR, along which the primordial fluctuations of the metric during inflation remain sufficiently small, as one can also see from equations (\ref{Pscalar-approx})-(\ref{Ptensor-approx}). 
Provided that at sufficiently high energy the dimensionless couplings $\widetilde{G}$ and $b$ start close to their fixed point value, 
given in (\ref{UVFP2}), and in particular provided that $b(k_0) = b_{0} \ll 1$, the RG evolution is stable, and connects 
smoothly the UV with the IR regime. At lower energies, with $\nicefrac{k}{m_{p}}\ll 1$ we have $\widetilde{G} \equiv k^2 G(k) \ll 1$ and 
$b \simeq b_{0} = \text{const.}$ Under these conditions, inflation can occur for a wide range of cut--off (or curvature) scales from the GUT scale down to the electroweak scale. The CMB data provides us with the appropriate renormalisation condition for the coupling $b$ at the energy scale where inflation occurs along the RG flow, relation (\ref{renorm-cond-b}), selecting out a particular RG trajectory among the infinitely many. 

\textbullet $\;$ We found evidence that the fixed point and associated RG dynamics we found for the Starobinsky action persists under the inclusion of a cubic term in the original effective action. Although this does not provide a conclusive proof, it provides encouraging evidence for its existence in higher order truncations. In particular, as we discussed earlier, under the appropriate renormalisation conditions, the RG flow will lead to a Planck--suppression of the cubic term, compared to the curvature squared one during the inflationary period, with the latter's coupling remaining sufficiently small and constant.

In the future, we plan to extend the analysis of this fascinating area \cite{CRS-in-prog}, presenting a more detailed exposition of 
the RG dynamics of the action (\ref{action:R^2}), from the UV to IR, together with a determination of further properties of the non-perturbative beta 
functions, including an explicit, further investigation into how the RG, inflationary dynamics and predictions  are  modified by 
the inclusion of higher order curvature operators in the action.

\acknowledgments
We are thankful to Mark Hindmarsh, Kirill Krasnov, Andrei Linde, Daniel Litim, Tony Padilla, Paul Saffin, Ignacy Sawicki, for useful discussions, and feedback. We also thank the referees for constructive feedback.
EJC acknowledges support by STFC and the Leverhulme Trust. IDS was supported by STFC. 

\appendix

\section{Explicit beta functions and the UVFP2}\label{App:Explicit-eqs}

In this section, we present more details about the actual caclulation of the RG flow, from which we derive explicitly the beta functions (\ref{betas_full}) for the couplings $\widetilde{G}$ and $b$ of the Starobinsky action used in this work. We will not present explicitly the beta functions associated with the $R^3$ truncation, as that would require a large amount of space. However, it is a straightforward task to extract them from the flow equation of Ref. (\cite{Codello:2008vh}). We will also comment on the nature of the non-trivial anomalous dimension of $b$ in the UV. We leave a more detailed discussion on its nature to future work \cite{CRS-in-prog}.

As explained in section \ref{sec: betas}, using the background field method, the metric field is decomposed into a background and fluctuating piece as
\be
g_{\alpha \beta} = \bar{g}_{\alpha \beta} + \delta g_{\alpha \beta},
\ee
with the background spacetime assumed to be a four-dimensional Euclidean sphere. In this background, the fluctuating piece of the metric is decomposed in its irreducible components as
\be
\delta g_{\alpha \beta} = h^{TT}_{\alpha \beta} + 2\nabla_{(\alpha} \xi_{\beta)} + (\nabla_{\alpha} \nabla_{\beta}  - \frac{1}{4}\nabla^2 g_{\alpha \beta} )\sigma + \frac{1}{4}g_{\alpha \beta}h, \label{metric-decomp}
\ee
 corresponding to the transverse--traceless, vector and scalar parts respectively. Derivatives are defined with respect to the background metric. Given the above decomposition the full propagator becomes a sum of the different sectors in (\ref{metric-decomp}). What is more, the trace in (\ref{WetterichEq}) is regularised by modifying the different propagator entries through an appropriate scale--dependent regulator as ${\bf \Gamma}(-\nabla^2) \to {\bf \Gamma}^{(2)}(-\nabla^2) + {\bf R}_k(-\nabla^2)$, with boldface denoting the possible matrix structure. The regulator function is chosen to be the ``optimised" cut--off of Ref. \cite{Litim_Opt-CO1}, defined as ${R}_k(-\nabla^2) = (k^2 - (-\nabla^2))\Theta(k^2 - (-\nabla^2))$, while the gauge is the Landau type gauge adopted in \cite{Codello:2008vh}. 

With these assumptions and definitions the full, non-perturbative beta functions for the dimensionless couplings $\widetilde{G} \equiv G/k^2$ and $b$ of the Starobinsky action take the form
\begin{eqnarray}
k\frac{ d }{d k} \widetilde{G}  &=& \frac{A_0 (1-B_2)+A_2 B_0}{1-A_1-B_2+A_1 B_2-A_2 B_1}, \label{betag_coeff}\\
k\frac{ d }{d k} b                   &=& \frac{B_0 (1-A_1)+A_0 B_1}{1-A_1-B_2+A_1 B_2-A_2 B_1}\ , \label{betab_coeff}
\end{eqnarray}
where the coefficients $A_i$ and $B_i$ are rational functions of $\widetilde{G}$ and $b$, and are given below, in (\ref{App:beta-coeffA0})--(\ref{App:beta-coeffB2}).

The fixed points of the flow equations result from the roots of the numerators as long as they are not roots of the denominator, too. In particular, 
equation (\ref{betag_coeff}) requires for a fixed point
\begin{equation}
 A_0=-\frac{A_2 B_0}{1-B_2}\ .
\end{equation}
Inserting the above relation into the beta function (\ref{betab_coeff}) gives
\begin{equation}
k\frac{ d }{d k} b = \frac{B_0}{1-B_2}=-2b + {\cal O}(b^2)
\end{equation}
from which one sees that one of the eigenvalues of the stability matrix at the fixed point $b=0$ is $-2$.

The origin of this particular eigenvalue can be understood by studying the structure of the flow equation in the $f(R)$ ansatz. In particular, it turns out that this eigenvalue is directly related to those terms in the right hand side of the flow equation (\ref{Wett_RG-Eq}) which are projected on the $R^2$ term and which have a prefactor of $1/b$. 
Starting from the explicit expression for the RG flow of $f(R)$-theories, as it is presented in Ref. \cite{Codello:2008vh}, it can be seen that
for the case of our action ansatz (\ref{action:R^2}), after expanding in curvature $\tilde R$ (see definitions after eq. (\ref{Wett_RG-Eq})) the only suitable term occurs in the tensor part, but not the scalar part obtained after the standard SO(5) (York) decomposition of the metric fluctuation \cite{York:1973ia}. 
A main difference between the scalar and tensor contribution in the flow equation is the fact that the scalar part has, due to higher-derivative terms not affecting the tensor part, an additional overall factor of $b$ after expansion in $\tilde R$. Therefore, the lowest order contribution in $b$ to the beta function $\beta_b$ comes entirely from the tensor part. In particular, it originates from a term proportional to 
$k\partial_k f_{\tilde R}-2\tilde R f_{\tilde{R}\tilde{R}}$ giving this way the required contributions to $B_2$ and $B_0$ respectively. The second term results from the quadratic mass dimension of the 
Ricci scalar and is responsible for the eigenvalue $-2$. It is remarkable that this eigenvalue results entirely from the gauge-invariant tensor part and not the 
gauge-dependent scalar part. We plan to present a detailed explanation of this point  in \cite{CRS-in-prog}.

\begin{widetext}
\subsection{Beta function coefficients}

The explicit form of the coefficients  in eqs. (\ref{betag_coeff})--(\ref{betab_coeff}), 
is given by
\begin{eqnarray}
A_0&=&\frac{\widetilde{G} \left(b^3 (144 \pi -301 \widetilde{G})+3456 \pi  b^2 (17 \widetilde{G}-8 \pi ) \widetilde{G}+9216 \pi ^2
   b (144 \pi -323 \widetilde{G}) \widetilde{G}^2+17694720 \pi ^3 \widetilde{G}^4  \right)}{72 \pi  b (b-96 \pi 
   \widetilde{G})^2} \label{App:beta-coeffA0}
   \\
A_1&=&\frac{4 \widetilde{G} \left(b^3-225 \pi  b^2 \widetilde{G}+15840 \pi ^2 b \widetilde{G}^2-276480 \pi ^3
   \widetilde{G}^3\right)}{9 \pi  b (b-96 \pi  \widetilde{G})^2} \label{App:beta-coeffA1}
\\
A_2&=&
\frac{16 \widetilde{G}^3 \left(b^2-200 \pi  b \widetilde{G}+7680 \pi ^2 \widetilde{G}^2\right)}{b^2 (b-96 \pi 
   \widetilde{G})^2}  \label{App:beta-coeffA2}
\\
B_0&=&
-\frac{491 b^5-157088 \pi  b^4 \widetilde{G}+18275328 \pi ^2 b^3 \widetilde{G}^2-916586496 \pi ^3 b^2
   \widetilde{G}^3+17694720000 \pi ^4 b \widetilde{G}^4-135895449600 \pi ^5 \widetilde{G}^5}{2880 \pi ^2 (b-96 \pi
    \widetilde{G})^3} \label{App:beta-coeffB0}
\\
B_1&=&
-\frac{-89 b^5+31818 \pi  b^4 \widetilde{G}-4328064 \pi ^2 b^3 \widetilde{G}^2+276203520 \pi ^3 b^2
   \widetilde{G}^3-8493465600 \pi ^4 b \widetilde{G}^4+101921587200 \pi ^5 \widetilde{G}^5}{4320 \pi ^2 \widetilde{G} (b-96
   \pi  \widetilde{G})^3} \label{App:beta-coeffB1}
\\
B_2&=&
\frac{\widetilde{G} \left(731 b^4-222912 \pi  b^3 \widetilde{G}+24247296 \pi ^2 b^2 \widetilde{G}^2-1150156800
   \pi ^3 b \widetilde{G}^3+16986931200 \pi ^4 \widetilde{G}^4 \right)}{720 \pi  b (b-96 \pi  \widetilde{G})^3} \label{App:beta-coeffB2}
\end{eqnarray}
\end{widetext}

\section{Slow--roll parameters and spectra}

\vspace{-0.15cm}

The slow--roll parameter $\epsilon \equiv - \nicefrac{\dot{H}}{H^2}$ was introduced in (\ref{slow-roll1}), and here we introduce the second--order slow--roll parameters as \cite{Hwang200113}
\begin{align}
& \epsilon_2 \equiv \frac{\dot{f_{R}}}{2H f_{R}} = \frac{f_{RR}}{2H f_{R}}\dot{R}\ , \\
& \epsilon_3 \equiv \frac{\ddot{f_{R}}}{H\dot{f_{R}}} = \frac{f_{RRR}}{f_{RR}} \frac{\dot{R}}{H} + \frac{\ddot{R}}{H\dot{R}}\ .
\end{align}

During slow-roll we have 
\be
R = 6H^2(2 - \epsilon), \; \dot{R}\simeq-24H^{3}\epsilon, \; \ddot{R} \simeq 24H^{4}\epsilon(3\epsilon - \nicefrac{\dot{\epsilon}}{H\epsilon}), 
\ee
$\nicefrac{\dot{\epsilon}}{H\epsilon} \simeq 2\epsilon$, and for the action (\ref{action:R^2-RG-improved2}) one has
$f_{RRR} = 0$ and $\nicefrac{\ddot{H}}{H^2}\simeq 0$. We can then approximate
\be
\epsilon_{2} \simeq  -\epsilon, \; \; \epsilon_{3} = \frac{\ddot{R}}{H\dot{R}} \simeq -\epsilon. \label{epsilon-approximations}
\ee

The tree-level scalar and tensor power spectrum respectively can be found to be 
\cite{Mukhanov:1990me,PhysRevD.71.063536,PhysRevD.54.1460,DeFelice:2010aj} (we follow the definitions in \cite{DeFelice:2010aj})
\begin{align}
& P_S = \frac{(1+\epsilon_2)^2}{12 f_{R}} \frac{1}{\epsilon_{3}^2} \left[ (1-\epsilon)\frac{\Gamma(2-\nicefrac{n_s}{2})}{\Gamma(3/2)} \frac{H}{2 \pi} \right]^2 \left( \frac{1}{2} | k_{F} \eta| \right)^{n_S - 1}, \label{Pscalar-general} \\
& P_T \simeq\frac{H^2}{\pi^2 f_{R}} \left( (1-\epsilon)\frac{\Gamma(\frac{3-n_T}{2})}{\Gamma(3/2)} \right)^2 \left( \frac{1}{2} | k_{F} \eta| \right)^{n_T}, \label{Ptensor-general}
\end{align}
with $\eta \equiv \int a^{-1} dt$ the conformal time, and the corresponding spectral indices defined as $n_S -1 \simeq -4\epsilon + 2\epsilon_2 - 2\epsilon_3$, and $n_T \simeq -2\epsilon - 2\epsilon_2$ respectively. 
Notice that the Fourier wave number is denoted by $k_{F}$ to distinguish it from 
the RG scale $k$. Using the slow-roll approximations of (\ref{epsilon-approximations}) we have that
\be
n_S -1 \simeq - 4 \epsilon, \; \; \; n_T \simeq 0. \label{nS}
\ee


\bibliography{AS_inflation.bib}

\begin{thebibliography}{107}%
\makeatletter
\providecommand \@ifxundefined [1]{%
 \@ifx{#1\undefined}
}%
\providecommand \@ifnum [1]{%
 \ifnum #1\expandafter \@firstoftwo
 \else \expandafter \@secondoftwo
 \fi
}%
\providecommand \@ifx [1]{%
 \ifx #1\expandafter \@firstoftwo
 \else \expandafter \@secondoftwo
 \fi
}%
\providecommand \natexlab [1]{#1}%
\providecommand \enquote  [1]{``#1''}%
\providecommand \bibnamefont  [1]{#1}%
\providecommand \bibfnamefont [1]{#1}%
\providecommand \citenamefont [1]{#1}%
\providecommand \href@noop [0]{\@secondoftwo}%
\providecommand \href [0]{\begingroup \@sanitize@url \@href}%
\providecommand \@href[1]{\@@startlink{#1}\@@href}%
\providecommand \@@href[1]{\endgroup#1\@@endlink}%
\providecommand \@sanitize@url [0]{\catcode `\\12\catcode `\$12\catcode
  `\&12\catcode `\#12\catcode `\^12\catcode `\_12\catcode `\%12\relax}%
\providecommand \@@startlink[1]{}%
\providecommand \@@endlink[0]{}%
\providecommand \url  [0]{\begingroup\@sanitize@url \@url }%
\providecommand \@url [1]{\endgroup\@href {#1}{\urlprefix }}%
\providecommand \urlprefix  [0]{URL }%
\providecommand \Eprint [0]{\href }%
\providecommand \doibase [0]{http://dx.doi.org/}%
\providecommand \selectlanguage [0]{\@gobble}%
\providecommand \bibinfo  [0]{\@secondoftwo}%
\providecommand \bibfield  [0]{\@secondoftwo}%
\providecommand \translation [1]{[#1]}%
\providecommand \BibitemOpen [0]{}%
\providecommand \bibitemStop [0]{}%
\providecommand \bibitemNoStop [0]{.\EOS\space}%
\providecommand \EOS [0]{\spacefactor3000\relax}%
\providecommand \BibitemShut  [1]{\csname bibitem#1\endcsname}%
\let\auto@bib@innerbib\@empty
\bibitem [{\citenamefont {Linde}(1984)}]{Linde:1984ir}%
  \BibitemOpen
  \bibfield  {author} {\bibinfo {author} {\bibfnamefont {A.~D.}\ \bibnamefont
  {Linde}},\ }\href {\doibase 10.1088/0034-4885/47/8/002} {\bibfield  {journal}
  {\bibinfo  {journal} {Rept.Prog.Phys.}\ }\textbf {\bibinfo {volume} {47}},\
  \bibinfo {pages} {925} (\bibinfo {year} {1984})}\BibitemShut {NoStop}%
\bibitem [{\citenamefont {Brandenberger}(1985)}]{Brandenberger:1984cz}%
  \BibitemOpen
  \bibfield  {author} {\bibinfo {author} {\bibfnamefont {R.~H.}\ \bibnamefont
  {Brandenberger}},\ }\href {\doibase 10.1103/RevModPhys.57.1} {\bibfield
  {journal} {\bibinfo  {journal} {Rev.Mod.Phys.}\ }\textbf {\bibinfo {volume}
  {57}},\ \bibinfo {pages} {1} (\bibinfo {year} {1985})}\BibitemShut {NoStop}%
\bibitem [{\citenamefont {Lyth}\ and\ \citenamefont
  {Riotto}(1999)}]{Lyth:1998xn}%
  \BibitemOpen
  \bibfield  {author} {\bibinfo {author} {\bibfnamefont {D.~H.}\ \bibnamefont
  {Lyth}}\ and\ \bibinfo {author} {\bibfnamefont {A.}~\bibnamefont {Riotto}},\
  }\href {\doibase 10.1016/S0370-1573(98)00128-8} {\bibfield  {journal}
  {\bibinfo  {journal} {Phys.Rept.}\ }\textbf {\bibinfo {volume} {314}},\
  \bibinfo {pages} {1} (\bibinfo {year} {1999})},\ \Eprint
  {http://arxiv.org/abs/hep-ph/9807278} {arXiv:hep-ph/9807278 [hep-ph]}
  \BibitemShut {NoStop}%
\bibitem [{\citenamefont {Lyth}\ and\ \citenamefont
  {Liddle}(2009)}]{Lyth:2009zz}%
  \BibitemOpen
  \bibfield  {author} {\bibinfo {author} {\bibfnamefont {D.~H.}\ \bibnamefont
  {Lyth}}\ and\ \bibinfo {author} {\bibfnamefont {A.~R.}\ \bibnamefont
  {Liddle}},\ }\href@noop {} {\  (\bibinfo {year} {2009})},\ \bibinfo {note}
  {{Cambridge University Press}}\BibitemShut {NoStop}%
\bibitem [{\citenamefont {Mazumdar}\ and\ \citenamefont
  {Rocher}(2011)}]{Mazumdar:2010sa}%
  \BibitemOpen
  \bibfield  {author} {\bibinfo {author} {\bibfnamefont {A.}~\bibnamefont
  {Mazumdar}}\ and\ \bibinfo {author} {\bibfnamefont {J.}~\bibnamefont
  {Rocher}},\ }\href {\doibase 10.1016/j.physrep.2010.08.001} {\bibfield
  {journal} {\bibinfo  {journal} {Phys.Rept.}\ }\textbf {\bibinfo {volume}
  {497}},\ \bibinfo {pages} {85} (\bibinfo {year} {2011})},\ \Eprint
  {http://arxiv.org/abs/1001.0993} {arXiv:1001.0993 [hep-ph]} \BibitemShut
  {NoStop}%
\bibitem [{\citenamefont {Starobinsky}(1980)}]{Star-R^2}%
  \BibitemOpen
  \bibfield  {author} {\bibinfo {author} {\bibfnamefont {A.~A.}\ \bibnamefont
  {Starobinsky}},\ }\href {\doibase 10.1016/0370-2693(80)90670-X} {\bibfield
  {journal} {\bibinfo  {journal} {Phys.Lett.}\ }\textbf {\bibinfo {volume}
  {B91}},\ \bibinfo {pages} {99} (\bibinfo {year} {1980})}\BibitemShut
  {NoStop}%
\bibitem [{\citenamefont {Ade}\ \emph {et~al.}(2015)\citenamefont {Ade} \emph
  {et~al.}}]{Ade:2015lrj}%
  \BibitemOpen
  \bibfield  {author} {\bibinfo {author} {\bibfnamefont {P.}~\bibnamefont
  {Ade}} \emph {et~al.} (\bibinfo {collaboration} {Planck}),\ }\href@noop {} {\
   (\bibinfo {year} {2015})},\ \Eprint {http://arxiv.org/abs/1502.02114}
  {arXiv:1502.02114 [astro-ph.CO]} \BibitemShut {NoStop}%
\bibitem [{\citenamefont {Vilenkin}(1985)}]{Vilenkin:1985md}%
  \BibitemOpen
  \bibfield  {author} {\bibinfo {author} {\bibfnamefont {A.}~\bibnamefont
  {Vilenkin}},\ }\href {\doibase 10.1103/PhysRevD.32.2511} {\bibfield
  {journal} {\bibinfo  {journal} {Phys.Rev.}\ }\textbf {\bibinfo {volume}
  {D32}},\ \bibinfo {pages} {2511} (\bibinfo {year} {1985})}\BibitemShut
  {NoStop}%
\bibitem [{\citenamefont {Birrell}\ and\ \citenamefont
  {Davies}(1982)}]{Birrell:1982ix}%
  \BibitemOpen
  \bibfield  {author} {\bibinfo {author} {\bibfnamefont {N.}~\bibnamefont
  {Birrell}}\ and\ \bibinfo {author} {\bibfnamefont {P.}~\bibnamefont
  {Davies}},\ }\href@noop {} {\emph {\bibinfo {title} {Quantum Fields in Curved
  Space}}}\ (\bibinfo  {publisher} {Cambridge Monogr.Math.Phys.},\ \bibinfo
  {year} {1982})\BibitemShut {NoStop}%
\bibitem [{\citenamefont {Ade}\ \emph {et~al.}(2013)\citenamefont {Ade} \emph
  {et~al.}}]{Ade:2013uln}%
  \BibitemOpen
  \bibfield  {author} {\bibinfo {author} {\bibfnamefont {P.}~\bibnamefont
  {Ade}} \emph {et~al.} (\bibinfo {collaboration} {Planck Collaboration}),\
  }\href@noop {} {\  (\bibinfo {year} {2013})},\ \Eprint
  {http://arxiv.org/abs/1303.5082} {arXiv:1303.5082 [astro-ph.CO]} \BibitemShut
  {NoStop}%
\bibitem [{\citenamefont {Dalianis}\ \emph {et~al.}(2015)\citenamefont
  {Dalianis}, \citenamefont {Farakos}, \citenamefont {Kehagias}, \citenamefont
  {Riotto},\ and\ \citenamefont {von Unge}}]{Dalianis:2014aya}%
  \BibitemOpen
  \bibfield  {author} {\bibinfo {author} {\bibfnamefont {I.}~\bibnamefont
  {Dalianis}}, \bibinfo {author} {\bibfnamefont {F.}~\bibnamefont {Farakos}},
  \bibinfo {author} {\bibfnamefont {A.}~\bibnamefont {Kehagias}}, \bibinfo
  {author} {\bibfnamefont {A.}~\bibnamefont {Riotto}}, \ and\ \bibinfo {author}
  {\bibfnamefont {R.}~\bibnamefont {von Unge}},\ }\href {\doibase
  10.1007/JHEP01(2015)043} {\bibfield  {journal} {\bibinfo  {journal} {JHEP}\
  }\textbf {\bibinfo {volume} {1501}},\ \bibinfo {pages} {043} (\bibinfo {year}
  {2015})},\ \Eprint {http://arxiv.org/abs/1409.8299} {arXiv:1409.8299
  [hep-th]} \BibitemShut {NoStop}%
\bibitem [{\citenamefont {Ferrara}\ \emph
  {et~al.}(2013{\natexlab{a}})\citenamefont {Ferrara}, \citenamefont {Kallosh},
  \citenamefont {Linde},\ and\ \citenamefont {Porrati}}]{Ferrara:2013rsa}%
  \BibitemOpen
  \bibfield  {author} {\bibinfo {author} {\bibfnamefont {S.}~\bibnamefont
  {Ferrara}}, \bibinfo {author} {\bibfnamefont {R.}~\bibnamefont {Kallosh}},
  \bibinfo {author} {\bibfnamefont {A.}~\bibnamefont {Linde}}, \ and\ \bibinfo
  {author} {\bibfnamefont {M.}~\bibnamefont {Porrati}},\ }\href@noop {} {\
  (\bibinfo {year} {2013}{\natexlab{a}})},\ \Eprint
  {http://arxiv.org/abs/1307.7696} {arXiv:1307.7696 [hep-th]} \BibitemShut
  {NoStop}%
\bibitem [{\citenamefont {Kallosh}\ and\ \citenamefont
  {Linde}(2013{\natexlab{a}})}]{Kallosh:2013hoa}%
  \BibitemOpen
  \bibfield  {author} {\bibinfo {author} {\bibfnamefont {R.}~\bibnamefont
  {Kallosh}}\ and\ \bibinfo {author} {\bibfnamefont {A.}~\bibnamefont
  {Linde}},\ }\href {\doibase 10.1088/1475-7516/2013/07/002} {\bibfield
  {journal} {\bibinfo  {journal} {JCAP}\ }\textbf {\bibinfo {volume} {1307}},\
  \bibinfo {pages} {002} (\bibinfo {year} {2013}{\natexlab{a}})},\ \Eprint
  {http://arxiv.org/abs/1306.5220} {arXiv:1306.5220 [hep-th]} \BibitemShut
  {NoStop}%
\bibitem [{\citenamefont {Kallosh}\ and\ \citenamefont
  {Linde}(2013{\natexlab{b}})}]{Kallosh:2013lkr}%
  \BibitemOpen
  \bibfield  {author} {\bibinfo {author} {\bibfnamefont {R.}~\bibnamefont
  {Kallosh}}\ and\ \bibinfo {author} {\bibfnamefont {A.}~\bibnamefont
  {Linde}},\ }\href {\doibase 10.1088/1475-7516/2013/06/028} {\bibfield
  {journal} {\bibinfo  {journal} {JCAP}\ }\textbf {\bibinfo {volume} {1306}},\
  \bibinfo {pages} {028} (\bibinfo {year} {2013}{\natexlab{b}})},\ \Eprint
  {http://arxiv.org/abs/1306.3214} {arXiv:1306.3214 [hep-th]} \BibitemShut
  {NoStop}%
\bibitem [{\citenamefont {Ferrara}\ \emph
  {et~al.}(2013{\natexlab{b}})\citenamefont {Ferrara}, \citenamefont {Kallosh},
  \citenamefont {Linde},\ and\ \citenamefont {Porrati}}]{Ferrara:2013kca}%
  \BibitemOpen
  \bibfield  {author} {\bibinfo {author} {\bibfnamefont {S.}~\bibnamefont
  {Ferrara}}, \bibinfo {author} {\bibfnamefont {R.}~\bibnamefont {Kallosh}},
  \bibinfo {author} {\bibfnamefont {A.}~\bibnamefont {Linde}}, \ and\ \bibinfo
  {author} {\bibfnamefont {M.}~\bibnamefont {Porrati}},\ }\href@noop {} {\
  (\bibinfo {year} {2013}{\natexlab{b}})},\ \Eprint
  {http://arxiv.org/abs/1309.1085} {arXiv:1309.1085 [hep-th]} \BibitemShut
  {NoStop}%
\bibitem [{\citenamefont {Ellis}\ \emph
  {et~al.}(2013{\natexlab{a}})\citenamefont {Ellis}, \citenamefont
  {Nanopoulos},\ and\ \citenamefont {Olive}}]{Ellis:2013xoa}%
  \BibitemOpen
  \bibfield  {author} {\bibinfo {author} {\bibfnamefont {J.}~\bibnamefont
  {Ellis}}, \bibinfo {author} {\bibfnamefont {D.~V.}\ \bibnamefont
  {Nanopoulos}}, \ and\ \bibinfo {author} {\bibfnamefont {K.~A.}\ \bibnamefont
  {Olive}},\ }\href {\doibase 10.1103/PhysRevLett.111.111301} {\bibfield
  {journal} {\bibinfo  {journal} {Phys.Rev.Lett.}\ }\textbf {\bibinfo {volume}
  {111}},\ \bibinfo {pages} {111301} (\bibinfo {year} {2013}{\natexlab{a}})},\
  \Eprint {http://arxiv.org/abs/1305.1247} {arXiv:1305.1247 [hep-th]}
  \BibitemShut {NoStop}%
\bibitem [{\citenamefont {Ellis}\ \emph
  {et~al.}(2013{\natexlab{b}})\citenamefont {Ellis}, \citenamefont
  {Nanopoulos},\ and\ \citenamefont {Olive}}]{Ellis:2013nxa}%
  \BibitemOpen
  \bibfield  {author} {\bibinfo {author} {\bibfnamefont {J.}~\bibnamefont
  {Ellis}}, \bibinfo {author} {\bibfnamefont {D.~V.}\ \bibnamefont
  {Nanopoulos}}, \ and\ \bibinfo {author} {\bibfnamefont {K.~A.}\ \bibnamefont
  {Olive}},\ }\href@noop {} {\  (\bibinfo {year} {2013}{\natexlab{b}})},\
  \Eprint {http://arxiv.org/abs/1307.3537} {arXiv:1307.3537 [hep-th]}
  \BibitemShut {NoStop}%
\bibitem [{\citenamefont {Farakos}\ \emph {et~al.}(2013)\citenamefont
  {Farakos}, \citenamefont {Kehagias},\ and\ \citenamefont
  {Riotto}}]{Farakos:2013cqa}%
  \BibitemOpen
  \bibfield  {author} {\bibinfo {author} {\bibfnamefont {F.}~\bibnamefont
  {Farakos}}, \bibinfo {author} {\bibfnamefont {A.}~\bibnamefont {Kehagias}}, \
  and\ \bibinfo {author} {\bibfnamefont {A.}~\bibnamefont {Riotto}},\ }\href
  {\doibase 10.1016/j.nuclphysb.2013.08.005} {\bibfield  {journal} {\bibinfo
  {journal} {Nucl.Phys.}\ }\textbf {\bibinfo {volume} {B876}},\ \bibinfo
  {pages} {187} (\bibinfo {year} {2013})},\ \Eprint
  {http://arxiv.org/abs/1307.1137} {arXiv:1307.1137 [hep-th]} \BibitemShut
  {NoStop}%
\bibitem [{\citenamefont {Alexandre}\ \emph {et~al.}(2014)\citenamefont
  {Alexandre}, \citenamefont {Houston},\ and\ \citenamefont
  {Mavromatos}}]{Alexandre:2013nqa}%
  \BibitemOpen
  \bibfield  {author} {\bibinfo {author} {\bibfnamefont {J.}~\bibnamefont
  {Alexandre}}, \bibinfo {author} {\bibfnamefont {N.}~\bibnamefont {Houston}},
  \ and\ \bibinfo {author} {\bibfnamefont {N.~E.}\ \bibnamefont {Mavromatos}},\
  }\href {\doibase 10.1103/PhysRevD.89.027703} {\bibfield  {journal} {\bibinfo
  {journal} {Phys.Rev.}\ }\textbf {\bibinfo {volume} {D89}},\ \bibinfo {pages}
  {027703} (\bibinfo {year} {2014})},\ \Eprint {http://arxiv.org/abs/1312.5197}
  {arXiv:1312.5197 [gr-qc]} \BibitemShut {NoStop}%
\bibitem [{\citenamefont {Alexandre}\ \emph {et~al.}(2015)\citenamefont
  {Alexandre}, \citenamefont {Houston},\ and\ \citenamefont
  {Mavromatos}}]{Alexandre:2014lla}%
  \BibitemOpen
  \bibfield  {author} {\bibinfo {author} {\bibfnamefont {J.}~\bibnamefont
  {Alexandre}}, \bibinfo {author} {\bibfnamefont {N.}~\bibnamefont {Houston}},
  \ and\ \bibinfo {author} {\bibfnamefont {N.~E.}\ \bibnamefont {Mavromatos}},\
  }\href {\doibase 10.1142/S0218271815410047} {\bibfield  {journal} {\bibinfo
  {journal} {Int.J.Mod.Phys.}\ }\textbf {\bibinfo {volume} {D24}},\ \bibinfo
  {pages} {1541004} (\bibinfo {year} {2015})},\ \Eprint
  {http://arxiv.org/abs/1409.3183} {arXiv:1409.3183 [gr-qc]} \BibitemShut
  {NoStop}%
\bibitem [{\citenamefont {Pallis}(2014)}]{Pallis:2013yda}%
  \BibitemOpen
  \bibfield  {author} {\bibinfo {author} {\bibfnamefont {C.}~\bibnamefont
  {Pallis}},\ }\href {\doibase 10.1088/1475-7516/2014/04/024} {\bibfield
  {journal} {\bibinfo  {journal} {JCAP}\ }\textbf {\bibinfo {volume} {1404}},\
  \bibinfo {pages} {024} (\bibinfo {year} {2014})},\ \Eprint
  {http://arxiv.org/abs/1312.3623} {arXiv:1312.3623 [hep-ph]} \BibitemShut
  {NoStop}%
\bibitem [{\citenamefont {Ellis}\ \emph {et~al.}(2014)\citenamefont {Ellis},
  \citenamefont {Mavromatos},\ and\ \citenamefont {Mulryne}}]{Ellis:2014rja}%
  \BibitemOpen
  \bibfield  {author} {\bibinfo {author} {\bibfnamefont {J.}~\bibnamefont
  {Ellis}}, \bibinfo {author} {\bibfnamefont {N.~E.}\ \bibnamefont
  {Mavromatos}}, \ and\ \bibinfo {author} {\bibfnamefont {D.~J.}\ \bibnamefont
  {Mulryne}},\ }\href {\doibase 10.1088/1475-7516/2014/05/012} {\bibfield
  {journal} {\bibinfo  {journal} {JCAP}\ }\textbf {\bibinfo {volume} {1405}},\
  \bibinfo {pages} {012} (\bibinfo {year} {2014})},\ \Eprint
  {http://arxiv.org/abs/1401.6078} {arXiv:1401.6078 [astro-ph.CO]} \BibitemShut
  {NoStop}%
\bibitem [{\citenamefont {Basilakos}\ \emph {et~al.}(2015)\citenamefont
  {Basilakos}, \citenamefont {Mavromatos},\ and\ \citenamefont
  {Sola}}]{Basilakos:2015yoa}%
  \BibitemOpen
  \bibfield  {author} {\bibinfo {author} {\bibfnamefont {S.}~\bibnamefont
  {Basilakos}}, \bibinfo {author} {\bibfnamefont {N.~E.}\ \bibnamefont
  {Mavromatos}}, \ and\ \bibinfo {author} {\bibfnamefont {J.}~\bibnamefont
  {Sola}},\ }\href@noop {} {\  (\bibinfo {year} {2015})},\ \Eprint
  {http://arxiv.org/abs/1505.04434} {arXiv:1505.04434 [gr-qc]} \BibitemShut
  {NoStop}%
\bibitem [{\citenamefont {Ferrara}\ \emph
  {et~al.}(2013{\natexlab{c}})\citenamefont {Ferrara}, \citenamefont
  {Kallosh},\ and\ \citenamefont {Van~Proeyen}}]{Ferrara:2013wka}%
  \BibitemOpen
  \bibfield  {author} {\bibinfo {author} {\bibfnamefont {S.}~\bibnamefont
  {Ferrara}}, \bibinfo {author} {\bibfnamefont {R.}~\bibnamefont {Kallosh}}, \
  and\ \bibinfo {author} {\bibfnamefont {A.}~\bibnamefont {Van~Proeyen}},\
  }\href@noop {} {\  (\bibinfo {year} {2013}{\natexlab{c}})},\ \Eprint
  {http://arxiv.org/abs/1309.4052} {arXiv:1309.4052 [hep-th]} \BibitemShut
  {NoStop}%
\bibitem [{\citenamefont {Briscese}\ \emph
  {et~al.}(2013{\natexlab{a}})\citenamefont {Briscese}, \citenamefont
  {Marciano}, \citenamefont {Modesto},\ and\ \citenamefont
  {Saridakis}}]{Briscese:2012ys}%
  \BibitemOpen
  \bibfield  {author} {\bibinfo {author} {\bibfnamefont {F.}~\bibnamefont
  {Briscese}}, \bibinfo {author} {\bibfnamefont {A.}~\bibnamefont {Marciano}},
  \bibinfo {author} {\bibfnamefont {L.}~\bibnamefont {Modesto}}, \ and\
  \bibinfo {author} {\bibfnamefont {E.~N.}\ \bibnamefont {Saridakis}},\ }\href
  {\doibase 10.1103/PhysRevD.87.083507} {\bibfield  {journal} {\bibinfo
  {journal} {Phys.Rev.}\ }\textbf {\bibinfo {volume} {D87}},\ \bibinfo {pages}
  {083507} (\bibinfo {year} {2013}{\natexlab{a}})},\ \Eprint
  {http://arxiv.org/abs/1212.3611} {arXiv:1212.3611 [hep-th]} \BibitemShut
  {NoStop}%
\bibitem [{\citenamefont {Kiritsis}(2013)}]{Kiritsis:2013gia}%
  \BibitemOpen
  \bibfield  {author} {\bibinfo {author} {\bibfnamefont {E.}~\bibnamefont
  {Kiritsis}},\ }\href@noop {} {\  (\bibinfo {year} {2013})},\ \Eprint
  {http://arxiv.org/abs/1307.5873} {arXiv:1307.5873 [hep-th]} \BibitemShut
  {NoStop}%
\bibitem [{\citenamefont {Briscese}\ \emph
  {et~al.}(2013{\natexlab{b}})\citenamefont {Briscese}, \citenamefont
  {Modesto},\ and\ \citenamefont {Tsujikawa}}]{Briscese:2013lna}%
  \BibitemOpen
  \bibfield  {author} {\bibinfo {author} {\bibfnamefont {F.}~\bibnamefont
  {Briscese}}, \bibinfo {author} {\bibfnamefont {L.}~\bibnamefont {Modesto}}, \
  and\ \bibinfo {author} {\bibfnamefont {S.}~\bibnamefont {Tsujikawa}},\
  }\href@noop {} {\  (\bibinfo {year} {2013}{\natexlab{b}})},\ \Eprint
  {http://arxiv.org/abs/1308.1413} {arXiv:1308.1413 [hep-th]} \BibitemShut
  {NoStop}%
\bibitem [{\citenamefont {Bamba}\ \emph
  {et~al.}(2014{\natexlab{a}})\citenamefont {Bamba}, \citenamefont {Cognola},
  \citenamefont {Odintsov},\ and\ \citenamefont {Zerbini}}]{Bamba:2014mua}%
  \BibitemOpen
  \bibfield  {author} {\bibinfo {author} {\bibfnamefont {K.}~\bibnamefont
  {Bamba}}, \bibinfo {author} {\bibfnamefont {G.}~\bibnamefont {Cognola}},
  \bibinfo {author} {\bibfnamefont {S.~D.}\ \bibnamefont {Odintsov}}, \ and\
  \bibinfo {author} {\bibfnamefont {S.}~\bibnamefont {Zerbini}},\ }\href
  {\doibase 10.1103/PhysRevD.90.023525} {\bibfield  {journal} {\bibinfo
  {journal} {Phys.Rev.}\ }\textbf {\bibinfo {volume} {D90}},\ \bibinfo {pages}
  {023525} (\bibinfo {year} {2014}{\natexlab{a}})},\ \Eprint
  {http://arxiv.org/abs/1404.4311} {arXiv:1404.4311 [gr-qc]} \BibitemShut
  {NoStop}%
\bibitem [{\citenamefont {Bamba}\ \emph
  {et~al.}(2014{\natexlab{b}})\citenamefont {Bamba}, \citenamefont
  {Myrzakulov}, \citenamefont {Odintsov},\ and\ \citenamefont
  {Sebastiani}}]{Bamba:2014jia}%
  \BibitemOpen
  \bibfield  {author} {\bibinfo {author} {\bibfnamefont {K.}~\bibnamefont
  {Bamba}}, \bibinfo {author} {\bibfnamefont {R.}~\bibnamefont {Myrzakulov}},
  \bibinfo {author} {\bibfnamefont {S.}~\bibnamefont {Odintsov}}, \ and\
  \bibinfo {author} {\bibfnamefont {L.}~\bibnamefont {Sebastiani}},\ }\href
  {\doibase 10.1103/PhysRevD.90.043505} {\bibfield  {journal} {\bibinfo
  {journal} {Phys.Rev.}\ }\textbf {\bibinfo {volume} {D90}},\ \bibinfo {pages}
  {043505} (\bibinfo {year} {2014}{\natexlab{b}})},\ \Eprint
  {http://arxiv.org/abs/1403.6649} {arXiv:1403.6649 [hep-th]} \BibitemShut
  {NoStop}%
\bibitem [{\citenamefont {Amorós}\ \emph {et~al.}(2014)\citenamefont
  {Amorós}, \citenamefont {de~Haro},\ and\ \citenamefont
  {Odintsov}}]{Amoros:2014tha}%
  \BibitemOpen
  \bibfield  {author} {\bibinfo {author} {\bibfnamefont {J.}~\bibnamefont
  {Amorós}}, \bibinfo {author} {\bibfnamefont {J.}~\bibnamefont {de~Haro}}, \
  and\ \bibinfo {author} {\bibfnamefont {S.~D.}\ \bibnamefont {Odintsov}},\
  }\href {\doibase 10.1103/PhysRevD.89.104010} {\bibfield  {journal} {\bibinfo
  {journal} {Phys.Rev.}\ }\textbf {\bibinfo {volume} {D89}},\ \bibinfo {pages}
  {104010} (\bibinfo {year} {2014})},\ \Eprint {http://arxiv.org/abs/1402.3071}
  {arXiv:1402.3071 [gr-qc]} \BibitemShut {NoStop}%
\bibitem [{\citenamefont {Linde}(1990)}]{Linde:2005ht}%
  \BibitemOpen
  \bibfield  {author} {\bibinfo {author} {\bibfnamefont {A.~D.}\ \bibnamefont
  {Linde}},\ }\href@noop {} {\bibfield  {journal} {\bibinfo  {journal}
  {Contemp.Concepts Phys.}\ }\textbf {\bibinfo {volume} {5}},\ \bibinfo {pages}
  {1} (\bibinfo {year} {1990})},\ \Eprint {http://arxiv.org/abs/hep-th/0503203}
  {arXiv:hep-th/0503203} \BibitemShut {NoStop}%
\bibitem [{\citenamefont {Linde}(1982)}]{Linde:1981mu}%
  \BibitemOpen
  \bibfield  {author} {\bibinfo {author} {\bibfnamefont {A.~D.}\ \bibnamefont
  {Linde}},\ }\href {\doibase 10.1016/0370-2693(82)91219-9} {\bibfield
  {journal} {\bibinfo  {journal} {Phys.Lett.}\ }\textbf {\bibinfo {volume}
  {B108}},\ \bibinfo {pages} {389} (\bibinfo {year} {1982})}\BibitemShut
  {NoStop}%
\bibitem [{\citenamefont {Kofman}\ \emph {et~al.}(1997)\citenamefont {Kofman},
  \citenamefont {Linde},\ and\ \citenamefont {Starobinsky}}]{Kofman:1997yn}%
  \BibitemOpen
  \bibfield  {author} {\bibinfo {author} {\bibfnamefont {L.}~\bibnamefont
  {Kofman}}, \bibinfo {author} {\bibfnamefont {A.~D.}\ \bibnamefont {Linde}}, \
  and\ \bibinfo {author} {\bibfnamefont {A.~A.}\ \bibnamefont {Starobinsky}},\
  }\href {\doibase 10.1103/PhysRevD.56.3258} {\bibfield  {journal} {\bibinfo
  {journal} {Phys.Rev.}\ }\textbf {\bibinfo {volume} {D56}},\ \bibinfo {pages}
  {3258} (\bibinfo {year} {1997})},\ \Eprint
  {http://arxiv.org/abs/hep-ph/9704452} {arXiv:hep-ph/9704452} \BibitemShut
  {NoStop}%
\bibitem [{\citenamefont {Copeland}\ \emph {et~al.}()\citenamefont {Copeland},
  \citenamefont {Rahmede},\ and\ \citenamefont {Saltas}}]{CRS-in-prog}%
  \BibitemOpen
  \bibfield  {author} {\bibinfo {author} {\bibfnamefont {E.}~\bibnamefont
  {Copeland}}, \bibinfo {author} {\bibfnamefont {C.}~\bibnamefont {Rahmede}}, \
  and\ \bibinfo {author} {\bibfnamefont {I.~D.}\ \bibnamefont {Saltas}},\
  }\href@noop {} {\ }\Eprint {http://arxiv.org/abs/In preparation} {In
  preparation} \BibitemShut {NoStop}%
\bibitem [{\citenamefont {Weinberg}(1979)}]{Weinberg:1980gg}%
  \BibitemOpen
  \bibfield  {author} {\bibinfo {author} {\bibfnamefont {S.}~\bibnamefont
  {Weinberg}},\ }\href@noop {} {\  (\bibinfo {year} {1979})},\ \bibinfo {note}
  {in General Relativity: An Einstein centenary survey, ed. S.~W. Hawking and
  W. Israel, 790- 831}\BibitemShut {NoStop}%
\bibitem [{\citenamefont {Reuter}(1998)}]{Reuter:1996cp}%
  \BibitemOpen
  \bibfield  {author} {\bibinfo {author} {\bibfnamefont {M.}~\bibnamefont
  {Reuter}},\ }\href {\doibase 10.1103/PhysRevD.57.971} {\bibfield  {journal}
  {\bibinfo  {journal} {Phys. Rev.}\ }\textbf {\bibinfo {volume} {D57}},\
  \bibinfo {pages} {971} (\bibinfo {year} {1998})},\ \Eprint
  {http://arxiv.org/abs/hep-th/9605030} {arXiv:hep-th/9605030} \BibitemShut
  {NoStop}%
\bibitem [{\citenamefont {Souma}(1999)}]{Souma:1999at}%
  \BibitemOpen
  \bibfield  {author} {\bibinfo {author} {\bibfnamefont {W.}~\bibnamefont
  {Souma}},\ }\href {\doibase 10.1143/PTP.102.181} {\bibfield  {journal}
  {\bibinfo  {journal} {Prog. Theor. Phys.}\ }\textbf {\bibinfo {volume}
  {102}},\ \bibinfo {pages} {181} (\bibinfo {year} {1999})},\ \Eprint
  {http://arxiv.org/abs/hep-th/9907027} {arXiv:hep-th/9907027} \BibitemShut
  {NoStop}%
\bibitem [{\citenamefont {Lauscher}\ and\ \citenamefont
  {Reuter}(2002{\natexlab{a}})}]{Lauscher:2001ya}%
  \BibitemOpen
  \bibfield  {author} {\bibinfo {author} {\bibfnamefont {O.}~\bibnamefont
  {Lauscher}}\ and\ \bibinfo {author} {\bibfnamefont {M.}~\bibnamefont
  {Reuter}},\ }\href {\doibase 10.1103/PhysRevD.65.025013} {\bibfield
  {journal} {\bibinfo  {journal} {Phys. Rev.}\ }\textbf {\bibinfo {volume}
  {D65}},\ \bibinfo {pages} {025013} (\bibinfo {year} {2002}{\natexlab{a}})},\
  \Eprint {http://arxiv.org/abs/hep-th/0108040} {arXiv:hep-th/0108040}
  \BibitemShut {NoStop}%
\bibitem [{\citenamefont {Lauscher}\ and\ \citenamefont
  {Reuter}(2002{\natexlab{b}})}]{Lauscher:2002sq}%
  \BibitemOpen
  \bibfield  {author} {\bibinfo {author} {\bibfnamefont {O.}~\bibnamefont
  {Lauscher}}\ and\ \bibinfo {author} {\bibfnamefont {M.}~\bibnamefont
  {Reuter}},\ }\href {\doibase 10.1103/PhysRevD.66.025026} {\bibfield
  {journal} {\bibinfo  {journal} {Phys. Rev.}\ }\textbf {\bibinfo {volume}
  {D66}},\ \bibinfo {pages} {025026} (\bibinfo {year} {2002}{\natexlab{b}})},\
  \Eprint {http://arxiv.org/abs/hep-th/0205062} {arXiv:hep-th/0205062}
  \BibitemShut {NoStop}%
\bibitem [{\citenamefont {Litim}(2004)}]{Litim:2003vp}%
  \BibitemOpen
  \bibfield  {author} {\bibinfo {author} {\bibfnamefont {D.~F.}\ \bibnamefont
  {Litim}},\ }\href {\doibase 10.1103/PhysRevLett.92.201301} {\bibfield
  {journal} {\bibinfo  {journal} {Phys. Rev. Lett.}\ }\textbf {\bibinfo
  {volume} {92}},\ \bibinfo {pages} {201301} (\bibinfo {year} {2004})},\
  \Eprint {http://arxiv.org/abs/hep-th/0312114} {arXiv:hep-th/0312114}
  \BibitemShut {NoStop}%
\bibitem [{\citenamefont {Reuter}\ and\ \citenamefont
  {Saueressig}(2002)}]{Reuter:2001ag}%
  \BibitemOpen
  \bibfield  {author} {\bibinfo {author} {\bibfnamefont {M.}~\bibnamefont
  {Reuter}}\ and\ \bibinfo {author} {\bibfnamefont {F.}~\bibnamefont
  {Saueressig}},\ }\href {\doibase 10.1103/PhysRevD.65.065016} {\bibfield
  {journal} {\bibinfo  {journal} {Phys. Rev.}\ }\textbf {\bibinfo {volume}
  {D65}},\ \bibinfo {pages} {065016} (\bibinfo {year} {2002})},\ \Eprint
  {http://arxiv.org/abs/hep-th/0110054} {arXiv:hep-th/0110054} \BibitemShut
  {NoStop}%
\bibitem [{\citenamefont {Dou}\ and\ \citenamefont
  {Percacci}(1998)}]{Dou:1997fg}%
  \BibitemOpen
  \bibfield  {author} {\bibinfo {author} {\bibfnamefont {D.}~\bibnamefont
  {Dou}}\ and\ \bibinfo {author} {\bibfnamefont {R.}~\bibnamefont {Percacci}},\
  }\href {\doibase 10.1088/0264-9381/15/11/011} {\bibfield  {journal} {\bibinfo
   {journal} {Class. Quant. Grav.}\ }\textbf {\bibinfo {volume} {15}},\
  \bibinfo {pages} {3449} (\bibinfo {year} {1998})},\ \Eprint
  {http://arxiv.org/abs/hep-th/9707239} {arXiv:hep-th/9707239} \BibitemShut
  {NoStop}%
\bibitem [{\citenamefont {Codello}\ \emph {et~al.}(2008)\citenamefont
  {Codello}, \citenamefont {Percacci},\ and\ \citenamefont
  {Rahmede}}]{Codello:2007bd}%
  \BibitemOpen
  \bibfield  {author} {\bibinfo {author} {\bibfnamefont {A.}~\bibnamefont
  {Codello}}, \bibinfo {author} {\bibfnamefont {R.}~\bibnamefont {Percacci}}, \
  and\ \bibinfo {author} {\bibfnamefont {C.}~\bibnamefont {Rahmede}},\ }\href
  {\doibase 10.1142/S0217751X08038135} {\bibfield  {journal} {\bibinfo
  {journal} {Int. J. Mod. Phys.}\ }\textbf {\bibinfo {volume} {A23}},\ \bibinfo
  {pages} {143} (\bibinfo {year} {2008})},\ \Eprint
  {http://arxiv.org/abs/0705.1769} {arXiv:0705.1769 [hep-th]} \BibitemShut
  {NoStop}%
\bibitem [{\citenamefont {Codello}\ \emph {et~al.}(2009)\citenamefont
  {Codello}, \citenamefont {Percacci},\ and\ \citenamefont
  {Rahmede}}]{Codello:2008vh}%
  \BibitemOpen
  \bibfield  {author} {\bibinfo {author} {\bibfnamefont {A.}~\bibnamefont
  {Codello}}, \bibinfo {author} {\bibfnamefont {R.}~\bibnamefont {Percacci}}, \
  and\ \bibinfo {author} {\bibfnamefont {C.}~\bibnamefont {Rahmede}},\ }\href
  {\doibase 10.1016/j.aop.2008.08.008} {\bibfield  {journal} {\bibinfo
  {journal} {Annals Phys.}\ }\textbf {\bibinfo {volume} {324}},\ \bibinfo
  {pages} {414} (\bibinfo {year} {2009})},\ \Eprint
  {http://arxiv.org/abs/0805.2909} {arXiv:0805.2909 [hep-th]} \BibitemShut
  {NoStop}%
\bibitem [{\citenamefont {Machado}\ and\ \citenamefont
  {Saueressig}(2008)}]{Machado:2007ea}%
  \BibitemOpen
  \bibfield  {author} {\bibinfo {author} {\bibfnamefont {P.~F.}\ \bibnamefont
  {Machado}}\ and\ \bibinfo {author} {\bibfnamefont {F.}~\bibnamefont
  {Saueressig}},\ }\href {\doibase 10.1103/PhysRevD.77.124045} {\bibfield
  {journal} {\bibinfo  {journal} {Phys. Rev.}\ }\textbf {\bibinfo {volume}
  {D77}},\ \bibinfo {pages} {124045} (\bibinfo {year} {2008})},\ \Eprint
  {http://arxiv.org/abs/0712.0445} {arXiv:0712.0445 [hep-th]} \BibitemShut
  {NoStop}%
\bibitem [{\citenamefont {Benedetti}\ \emph {et~al.}(2010)\citenamefont
  {Benedetti}, \citenamefont {Machado},\ and\ \citenamefont
  {Saueressig}}]{Benedetti:2009gn}%
  \BibitemOpen
  \bibfield  {author} {\bibinfo {author} {\bibfnamefont {D.}~\bibnamefont
  {Benedetti}}, \bibinfo {author} {\bibfnamefont {P.~F.}\ \bibnamefont
  {Machado}}, \ and\ \bibinfo {author} {\bibfnamefont {F.}~\bibnamefont
  {Saueressig}},\ }\href {\doibase 10.1016/j.nuclphysb.2009.08.023} {\bibfield
  {journal} {\bibinfo  {journal} {Nucl.Phys.}\ }\textbf {\bibinfo {volume}
  {B824}},\ \bibinfo {pages} {168} (\bibinfo {year} {2010})},\ \bibinfo {note}
  {16 pages},\ \Eprint {http://arxiv.org/abs/0902.4630} {arXiv:0902.4630
  [hep-th]} \BibitemShut {NoStop}%
\bibitem [{\citenamefont {Benedetti}\ \emph {et~al.}(2009)\citenamefont
  {Benedetti}, \citenamefont {Machado},\ and\ \citenamefont
  {Saueressig}}]{Benedetti:2009rx}%
  \BibitemOpen
  \bibfield  {author} {\bibinfo {author} {\bibfnamefont {D.}~\bibnamefont
  {Benedetti}}, \bibinfo {author} {\bibfnamefont {P.~F.}\ \bibnamefont
  {Machado}}, \ and\ \bibinfo {author} {\bibfnamefont {F.}~\bibnamefont
  {Saueressig}},\ }\href {\doibase 10.1142/S0217732309031521} {\bibfield
  {journal} {\bibinfo  {journal} {Mod.Phys.Lett.}\ }\textbf {\bibinfo {volume}
  {A24}},\ \bibinfo {pages} {2233} (\bibinfo {year} {2009})},\ \bibinfo {note}
  {4 pages},\ \Eprint {http://arxiv.org/abs/0901.2984} {arXiv:0901.2984
  [hep-th]} \BibitemShut {NoStop}%
\bibitem [{\citenamefont {Benedetti}\ and\ \citenamefont
  {Caravelli}(2012)}]{BenedettiCaravelli}%
  \BibitemOpen
  \bibfield  {author} {\bibinfo {author} {\bibfnamefont {D.}~\bibnamefont
  {Benedetti}}\ and\ \bibinfo {author} {\bibfnamefont {F.}~\bibnamefont
  {Caravelli}},\ }\href@noop {} {\  (\bibinfo {year} {2012})},\ \bibinfo {note}
  {31 pages, 7 figures},\ \Eprint {http://arxiv.org/abs/1204.3541}
  {arXiv:1204.3541 [hep-th]} \BibitemShut {NoStop}%
\bibitem [{\citenamefont {Benedetti}(2012)}]{Benedetti:2011ct}%
  \BibitemOpen
  \bibfield  {author} {\bibinfo {author} {\bibfnamefont {D.}~\bibnamefont
  {Benedetti}},\ }\href {\doibase 10.1088/1367-2630/14/1/015005} {\bibfield
  {journal} {\bibinfo  {journal} {New J.Phys.}\ }\textbf {\bibinfo {volume}
  {14}},\ \bibinfo {pages} {015005} (\bibinfo {year} {2012})},\ \Eprint
  {http://arxiv.org/abs/1107.3110} {arXiv:1107.3110 [hep-th]} \BibitemShut
  {NoStop}%
\bibitem [{\citenamefont {Falls}\ \emph {et~al.}(2013)\citenamefont {Falls},
  \citenamefont {Litim}, \citenamefont {Nikolakopoulos},\ and\ \citenamefont
  {Rahmede}}]{Falls:2013bv}%
  \BibitemOpen
  \bibfield  {author} {\bibinfo {author} {\bibfnamefont {K.}~\bibnamefont
  {Falls}}, \bibinfo {author} {\bibfnamefont {D.}~\bibnamefont {Litim}},
  \bibinfo {author} {\bibfnamefont {K.}~\bibnamefont {Nikolakopoulos}}, \ and\
  \bibinfo {author} {\bibfnamefont {C.}~\bibnamefont {Rahmede}},\ }\href@noop
  {} {\  (\bibinfo {year} {2013})},\ \Eprint {http://arxiv.org/abs/1301.4191}
  {arXiv:1301.4191 [hep-th]} \BibitemShut {NoStop}%
\bibitem [{\citenamefont {Manrique}\ \emph {et~al.}(2011)\citenamefont
  {Manrique}, \citenamefont {Rechenberger},\ and\ \citenamefont
  {Saueressig}}]{Manrique:2011jc}%
  \BibitemOpen
  \bibfield  {author} {\bibinfo {author} {\bibfnamefont {E.}~\bibnamefont
  {Manrique}}, \bibinfo {author} {\bibfnamefont {S.}~\bibnamefont
  {Rechenberger}}, \ and\ \bibinfo {author} {\bibfnamefont {F.}~\bibnamefont
  {Saueressig}},\ }\href {\doibase 10.1103/PhysRevLett.106.251302} {\bibfield
  {journal} {\bibinfo  {journal} {Phys.Rev.Lett.}\ }\textbf {\bibinfo {volume}
  {106}},\ \bibinfo {pages} {251302} (\bibinfo {year} {2011})},\ \Eprint
  {http://arxiv.org/abs/1102.5012} {arXiv:1102.5012 [hep-th]} \BibitemShut
  {NoStop}%
\bibitem [{\citenamefont {Eichhorn}(2013)}]{Eichhorn:2013ug}%
  \BibitemOpen
  \bibfield  {author} {\bibinfo {author} {\bibfnamefont {A.}~\bibnamefont
  {Eichhorn}},\ }\href {\doibase 10.1103/PhysRevD.87.124016} {\bibfield
  {journal} {\bibinfo  {journal} {Phys.Rev.}\ }\textbf {\bibinfo {volume}
  {D87}},\ \bibinfo {pages} {124016} (\bibinfo {year} {2013})},\ \Eprint
  {http://arxiv.org/abs/1301.0632} {arXiv:1301.0632 [hep-th]} \BibitemShut
  {NoStop}%
\bibitem [{\citenamefont {Demmel}\ \emph {et~al.}(2013)\citenamefont {Demmel},
  \citenamefont {Saueressig},\ and\ \citenamefont {Zanusso}}]{Demmel:2013myx}%
  \BibitemOpen
  \bibfield  {author} {\bibinfo {author} {\bibfnamefont {M.}~\bibnamefont
  {Demmel}}, \bibinfo {author} {\bibfnamefont {F.}~\bibnamefont {Saueressig}},
  \ and\ \bibinfo {author} {\bibfnamefont {O.}~\bibnamefont {Zanusso}},\
  }\href@noop {} {\  (\bibinfo {year} {2013})},\ \Eprint
  {http://arxiv.org/abs/1302.1312} {arXiv:1302.1312 [hep-th]} \BibitemShut
  {NoStop}%
\bibitem [{\citenamefont {Dietz}\ and\ \citenamefont
  {Morris}(2013{\natexlab{a}})}]{Dietz:2012ic}%
  \BibitemOpen
  \bibfield  {author} {\bibinfo {author} {\bibfnamefont {J.~A.}\ \bibnamefont
  {Dietz}}\ and\ \bibinfo {author} {\bibfnamefont {T.~R.}\ \bibnamefont
  {Morris}},\ }\href {\doibase 10.1007/JHEP01(2013)108} {\bibfield  {journal}
  {\bibinfo  {journal} {JHEP}\ }\textbf {\bibinfo {volume} {1301}},\ \bibinfo
  {pages} {108} (\bibinfo {year} {2013}{\natexlab{a}})},\ \Eprint
  {http://arxiv.org/abs/1211.0955} {arXiv:1211.0955 [hep-th]} \BibitemShut
  {NoStop}%
\bibitem [{\citenamefont {Dietz}\ and\ \citenamefont
  {Morris}(2013{\natexlab{b}})}]{Dietz:2013sba}%
  \BibitemOpen
  \bibfield  {author} {\bibinfo {author} {\bibfnamefont {J.~A.}\ \bibnamefont
  {Dietz}}\ and\ \bibinfo {author} {\bibfnamefont {T.~R.}\ \bibnamefont
  {Morris}},\ }\href {\doibase 10.1007/JHEP07(2013)064} {\bibfield  {journal}
  {\bibinfo  {journal} {JHEP}\ }\textbf {\bibinfo {volume} {1307}},\ \bibinfo
  {pages} {064} (\bibinfo {year} {2013}{\natexlab{b}})},\ \Eprint
  {http://arxiv.org/abs/1306.1223} {arXiv:1306.1223 [hep-th]} \BibitemShut
  {NoStop}%
\bibitem [{\citenamefont {Rechenberger}\ and\ \citenamefont
  {Saueressig}(2012)}]{Rechenberger:2012pm}%
  \BibitemOpen
  \bibfield  {author} {\bibinfo {author} {\bibfnamefont {S.}~\bibnamefont
  {Rechenberger}}\ and\ \bibinfo {author} {\bibfnamefont {F.}~\bibnamefont
  {Saueressig}},\ }\href {\doibase 10.1103/PhysRevD.86.024018} {\bibfield
  {journal} {\bibinfo  {journal} {Phys.Rev.}\ }\textbf {\bibinfo {volume}
  {D86}},\ \bibinfo {pages} {024018} (\bibinfo {year} {2012})},\ \Eprint
  {http://arxiv.org/abs/1206.0657} {arXiv:1206.0657 [hep-th]} \BibitemShut
  {NoStop}%
\bibitem [{\citenamefont {Niedermaier}\ and\ \citenamefont
  {Reuter}(2006)}]{Niedermaier:2006wt}%
  \BibitemOpen
  \bibfield  {author} {\bibinfo {author} {\bibfnamefont {M.}~\bibnamefont
  {Niedermaier}}\ and\ \bibinfo {author} {\bibfnamefont {M.}~\bibnamefont
  {Reuter}},\ }\href@noop {} {\bibfield  {journal} {\bibinfo  {journal} {Living
  Rev. Rel.}\ }\textbf {\bibinfo {volume} {9}},\ \bibinfo {pages} {5} (\bibinfo
  {year} {2006})}\BibitemShut {NoStop}%
\bibitem [{\citenamefont {Percacci}(2007)}]{Percacci:2007sz}%
  \BibitemOpen
  \bibfield  {author} {\bibinfo {author} {\bibfnamefont {R.}~\bibnamefont
  {Percacci}},\ }\href@noop {} {\  (\bibinfo {year} {2007})},\ \bibinfo {note}
  {in ``Approaches to Quantum Gravity: Towards a New Understanding of Space,
  Time and Matter'' ed. D. Oriti, Cambridge University Press},\ \Eprint
  {http://arxiv.org/abs/0709.3851} {arXiv:0709.3851 [hep-th]} \BibitemShut
  {NoStop}%
\bibitem [{\citenamefont {Litim}(2008)}]{Litim:2008tt}%
  \BibitemOpen
  \bibfield  {author} {\bibinfo {author} {\bibfnamefont {D.~F.}\ \bibnamefont
  {Litim}},\ }\href@noop {} {\  (\bibinfo {year} {2008})},\ \Eprint
  {http://arxiv.org/abs/0810.3675} {arXiv:0810.3675 [hep-th]} \BibitemShut
  {NoStop}%
\bibitem [{\citenamefont {Reuter}\ and\ \citenamefont
  {Saueressig}(2012)}]{Reuter:2012id}%
  \BibitemOpen
  \bibfield  {author} {\bibinfo {author} {\bibfnamefont {M.}~\bibnamefont
  {Reuter}}\ and\ \bibinfo {author} {\bibfnamefont {F.}~\bibnamefont
  {Saueressig}},\ }\href {\doibase 10.1088/1367-2630/14/5/055022} {\bibfield
  {journal} {\bibinfo  {journal} {New J.Phys.}\ }\textbf {\bibinfo {volume}
  {14}},\ \bibinfo {pages} {055022} (\bibinfo {year} {2012})},\ \Eprint
  {http://arxiv.org/abs/1202.2274} {arXiv:1202.2274 [hep-th]} \BibitemShut
  {NoStop}%
\bibitem [{\citenamefont {Contillo}\ \emph {et~al.}(2012)\citenamefont
  {Contillo}, \citenamefont {Hindmarsh},\ and\ \citenamefont
  {Rahmede}}]{Contillo:2011ag}%
  \BibitemOpen
  \bibfield  {author} {\bibinfo {author} {\bibfnamefont {A.}~\bibnamefont
  {Contillo}}, \bibinfo {author} {\bibfnamefont {M.}~\bibnamefont {Hindmarsh}},
  \ and\ \bibinfo {author} {\bibfnamefont {C.}~\bibnamefont {Rahmede}},\ }\href
  {\doibase 10.1103/PhysRevD.85.043501} {\bibfield  {journal} {\bibinfo
  {journal} {Phys.Rev.}\ }\textbf {\bibinfo {volume} {D85}},\ \bibinfo {pages}
  {043501} (\bibinfo {year} {2012})},\ \Eprint {http://arxiv.org/abs/1108.0422}
  {arXiv:1108.0422 [gr-qc]} \BibitemShut {NoStop}%
\bibitem [{\citenamefont {Kaya}(2013)}]{Kaya:2013bga}%
  \BibitemOpen
  \bibfield  {author} {\bibinfo {author} {\bibfnamefont {A.}~\bibnamefont
  {Kaya}},\ }\href {\doibase 10.1103/PhysRevD.87.123501} {\bibfield  {journal}
  {\bibinfo  {journal} {Phys.Rev.}\ }\textbf {\bibinfo {volume} {D87}},\
  \bibinfo {pages} {123501} (\bibinfo {year} {2013})},\ \Eprint
  {http://arxiv.org/abs/1303.5459} {arXiv:1303.5459 [hep-th]} \BibitemShut
  {NoStop}%
\bibitem [{\citenamefont {Cai}\ and\ \citenamefont {Easson}(2012)}]{Cai_RG}%
  \BibitemOpen
  \bibfield  {author} {\bibinfo {author} {\bibfnamefont {Y.-F.}\ \bibnamefont
  {Cai}}\ and\ \bibinfo {author} {\bibfnamefont {D.~A.}\ \bibnamefont
  {Easson}},\ }\href@noop {} {\  (\bibinfo {year} {2012})},\ \Eprint
  {http://arxiv.org/abs/1202.1285} {arXiv:1202.1285 [hep-th]} \BibitemShut
  {NoStop}%
\bibitem [{\citenamefont {Cai}\ \emph {et~al.}(2013)\citenamefont {Cai},
  \citenamefont {Chang}, \citenamefont {Chen}, \citenamefont {Easson},\ and\
  \citenamefont {Qiu}}]{Cai:2013caa}%
  \BibitemOpen
  \bibfield  {author} {\bibinfo {author} {\bibfnamefont {Y.-F.}\ \bibnamefont
  {Cai}}, \bibinfo {author} {\bibfnamefont {Y.-C.}\ \bibnamefont {Chang}},
  \bibinfo {author} {\bibfnamefont {P.}~\bibnamefont {Chen}}, \bibinfo {author}
  {\bibfnamefont {D.~A.}\ \bibnamefont {Easson}}, \ and\ \bibinfo {author}
  {\bibfnamefont {T.}~\bibnamefont {Qiu}},\ }\href {\doibase
  10.1103/PhysRevD.88.083508} {\bibfield  {journal} {\bibinfo  {journal}
  {Phys.Rev.}\ }\textbf {\bibinfo {volume} {D88}},\ \bibinfo {pages} {083508}
  (\bibinfo {year} {2013})},\ \Eprint {http://arxiv.org/abs/1304.6938}
  {arXiv:1304.6938 [hep-th]} \BibitemShut {NoStop}%
\bibitem [{\citenamefont {Weinberg}(2010)}]{Wein}%
  \BibitemOpen
  \bibfield  {author} {\bibinfo {author} {\bibfnamefont {S.}~\bibnamefont
  {Weinberg}},\ }\href {\doibase 10.1103/PhysRevD.81.083535} {\bibfield
  {journal} {\bibinfo  {journal} {Phys.Rev.}\ }\textbf {\bibinfo {volume}
  {D81}},\ \bibinfo {pages} {083535} (\bibinfo {year} {2010})},\ \Eprint
  {http://arxiv.org/abs/0911.3165} {arXiv:0911.3165 [hep-th]} \BibitemShut
  {NoStop}%
\bibitem [{\citenamefont {Reuter}\ and\ \citenamefont
  {Saueressig}(2005)}]{Reuter:2005kb}%
  \BibitemOpen
  \bibfield  {author} {\bibinfo {author} {\bibfnamefont {M.}~\bibnamefont
  {Reuter}}\ and\ \bibinfo {author} {\bibfnamefont {F.}~\bibnamefont
  {Saueressig}},\ }\href {\doibase 10.1088/1475-7516/2005/09/012} {\bibfield
  {journal} {\bibinfo  {journal} {JCAP}\ }\textbf {\bibinfo {volume} {0509}},\
  \bibinfo {pages} {012} (\bibinfo {year} {2005})},\ \Eprint
  {http://arxiv.org/abs/hep-th/0507167} {arXiv:hep-th/0507167} \BibitemShut
  {NoStop}%
\bibitem [{\citenamefont {Hindmarsh}\ and\ \citenamefont
  {Saltas}(2012)}]{Hindmarsh:2012rc}%
  \BibitemOpen
  \bibfield  {author} {\bibinfo {author} {\bibfnamefont {M.}~\bibnamefont
  {Hindmarsh}}\ and\ \bibinfo {author} {\bibfnamefont {I.~D.}\ \bibnamefont
  {Saltas}},\ }\href@noop {} {\bibfield  {journal} {\bibinfo  {journal}
  {Phys.Rev.}\ }\textbf {\bibinfo {volume} {D86}},\ \bibinfo {pages} {064029}
  (\bibinfo {year} {2012})},\ \Eprint {http://arxiv.org/abs/1203.3957}
  {arXiv:1203.3957 [gr-qc]} \BibitemShut {NoStop}%
\bibitem [{\citenamefont {Cai}\ and\ \citenamefont
  {Easson}(2011)}]{Cai:2011kd}%
  \BibitemOpen
  \bibfield  {author} {\bibinfo {author} {\bibfnamefont {Y.-F.}\ \bibnamefont
  {Cai}}\ and\ \bibinfo {author} {\bibfnamefont {D.~A.}\ \bibnamefont
  {Easson}},\ }\href {\doibase 10.1103/PhysRevD.84.103502} {\bibfield
  {journal} {\bibinfo  {journal} {Phys.Rev.}\ }\textbf {\bibinfo {volume}
  {D84}},\ \bibinfo {pages} {103502} (\bibinfo {year} {2011})},\ \Eprint
  {http://arxiv.org/abs/1107.5815} {arXiv:1107.5815 [hep-th]} \BibitemShut
  {NoStop}%
\bibitem [{\citenamefont {Fradkin}\ and\ \citenamefont
  {Tseytlin}(1981)}]{Fradkin:1981hx}%
  \BibitemOpen
  \bibfield  {author} {\bibinfo {author} {\bibfnamefont {E.}~\bibnamefont
  {Fradkin}}\ and\ \bibinfo {author} {\bibfnamefont {A.~A.}\ \bibnamefont
  {Tseytlin}},\ }\href {\doibase 10.1016/0370-2693(81)90702-4} {\bibfield
  {journal} {\bibinfo  {journal} {Phys.Lett.}\ }\textbf {\bibinfo {volume}
  {B104}},\ \bibinfo {pages} {377} (\bibinfo {year} {1981})}\BibitemShut
  {NoStop}%
\bibitem [{\citenamefont {Fradkin}\ and\ \citenamefont
  {Tseytlin}(1982)}]{Fradkin:1981iu}%
  \BibitemOpen
  \bibfield  {author} {\bibinfo {author} {\bibfnamefont {E.}~\bibnamefont
  {Fradkin}}\ and\ \bibinfo {author} {\bibfnamefont {A.~A.}\ \bibnamefont
  {Tseytlin}},\ }\href {\doibase 10.1016/0550-3213(82)90444-8} {\bibfield
  {journal} {\bibinfo  {journal} {Nucl.Phys.}\ }\textbf {\bibinfo {volume}
  {B201}},\ \bibinfo {pages} {469} (\bibinfo {year} {1982})}\BibitemShut
  {NoStop}%
\bibitem [{\citenamefont {Codello}\ and\ \citenamefont
  {Percacci}(2006)}]{Codello:2006in}%
  \BibitemOpen
  \bibfield  {author} {\bibinfo {author} {\bibfnamefont {A.}~\bibnamefont
  {Codello}}\ and\ \bibinfo {author} {\bibfnamefont {R.}~\bibnamefont
  {Percacci}},\ }\href {\doibase 10.1103/PhysRevLett.97.221301} {\bibfield
  {journal} {\bibinfo  {journal} {Phys.Rev.Lett.}\ }\textbf {\bibinfo {volume}
  {97}},\ \bibinfo {pages} {221301} (\bibinfo {year} {2006})},\ \Eprint
  {http://arxiv.org/abs/hep-th/0607128} {arXiv:hep-th/0607128 [hep-th]}
  \BibitemShut {NoStop}%
\bibitem [{\citenamefont {Niedermaier}(2010)}]{Niedermaier:2010zz}%
  \BibitemOpen
  \bibfield  {author} {\bibinfo {author} {\bibfnamefont {M.}~\bibnamefont
  {Niedermaier}},\ }\href {\doibase 10.1016/j.nuclphysb.2010.01.016} {\bibfield
   {journal} {\bibinfo  {journal} {Nucl.Phys.}\ }\textbf {\bibinfo {volume}
  {B833}},\ \bibinfo {pages} {226} (\bibinfo {year} {2010})}\BibitemShut
  {NoStop}%
\bibitem [{\citenamefont {Gies}(2012)}]{Gies:2006wv}%
  \BibitemOpen
  \bibfield  {author} {\bibinfo {author} {\bibfnamefont {H.}~\bibnamefont
  {Gies}},\ }\href {\doibase 10.1007/978-3-642-27320-9_6} {\bibfield  {journal}
  {\bibinfo  {journal} {Lect.Notes Phys.}\ }\textbf {\bibinfo {volume} {852}},\
  \bibinfo {pages} {287} (\bibinfo {year} {2012})},\ \Eprint
  {http://arxiv.org/abs/hep-ph/0611146} {arXiv:hep-ph/0611146 [hep-ph]}
  \BibitemShut {NoStop}%
\bibitem [{\citenamefont {Pawlowski}(2007)}]{Pawlowski:2005xe}%
  \BibitemOpen
  \bibfield  {author} {\bibinfo {author} {\bibfnamefont {J.~M.}\ \bibnamefont
  {Pawlowski}},\ }\href {\doibase 10.1016/j.aop.2007.01.007} {\bibfield
  {journal} {\bibinfo  {journal} {Annals Phys.}\ }\textbf {\bibinfo {volume}
  {322}},\ \bibinfo {pages} {2831} (\bibinfo {year} {2007})},\ \Eprint
  {http://arxiv.org/abs/hep-th/0512261} {arXiv:hep-th/0512261 [hep-th]}
  \BibitemShut {NoStop}%
\bibitem [{\citenamefont {Wetterich}(1993)}]{Wett_RG-Eq}%
  \BibitemOpen
  \bibfield  {author} {\bibinfo {author} {\bibfnamefont {C.}~\bibnamefont
  {Wetterich}},\ }\href {\doibase 10.1016/0370-2693(93)90726-X} {\bibfield
  {journal} {\bibinfo  {journal} {Phys.Lett.}\ }\textbf {\bibinfo {volume}
  {B301}},\ \bibinfo {pages} {90} (\bibinfo {year} {1993})}\BibitemShut
  {NoStop}%
\bibitem [{\citenamefont {Morris}(1994)}]{Morris:1994ie}%
  \BibitemOpen
  \bibfield  {author} {\bibinfo {author} {\bibfnamefont {T.~R.}\ \bibnamefont
  {Morris}},\ }\href {\doibase 10.1016/0370-2693(94)90767-6} {\bibfield
  {journal} {\bibinfo  {journal} {Phys.Lett.}\ }\textbf {\bibinfo {volume}
  {B329}},\ \bibinfo {pages} {241} (\bibinfo {year} {1994})},\ \Eprint
  {http://arxiv.org/abs/hep-ph/9403340} {arXiv:hep-ph/9403340 [hep-ph]}
  \BibitemShut {NoStop}%
\bibitem [{\citenamefont {Litim}(2000{\natexlab{a}})}]{Litim_Opt-CO1}%
  \BibitemOpen
  \bibfield  {author} {\bibinfo {author} {\bibfnamefont {D.~F.}\ \bibnamefont
  {Litim}},\ }\href {\doibase 10.1016/S0370-2693(00)00748-6} {\bibfield
  {journal} {\bibinfo  {journal} {Phys.Lett.}\ }\textbf {\bibinfo {volume}
  {B486}},\ \bibinfo {pages} {92} (\bibinfo {year} {2000}{\natexlab{a}})},\
  \Eprint {http://arxiv.org/abs/hep-th/0005245} {arXiv:hep-th/0005245}
  \BibitemShut {NoStop}%
\bibitem [{\citenamefont {Higgs}(1959)}]{Higgs_Auxiliary_Scalars}%
  \BibitemOpen
  \bibfield  {author} {\bibinfo {author} {\bibfnamefont {P.}~\bibnamefont
  {Higgs}},\ }\href
  {http://www.scopus.com/inward/record.url?eid=2-s2.0-51649195681&partnerID=40&md5=620774638aa6e5ba553a93e0443c66ed}
  {\bibfield  {journal} {\bibinfo  {journal} {Il Nuovo Cimento Series 10}\
  }\textbf {\bibinfo {volume} {11}},\ \bibinfo {pages} {816} (\bibinfo {year}
  {1959})},\ \bibinfo {note} {cited By (since 1996) 47}\BibitemShut {NoStop}%
\bibitem [{\citenamefont {Bicknell}(1974)}]{BIcknell_Auxiliary_Scalars}%
  \BibitemOpen
  \bibfield  {author} {\bibinfo {author} {\bibfnamefont {G.~V.}\ \bibnamefont
  {Bicknell}},\ }\href {http://stacks.iop.org/0301-0015/7/i=9/a=010} {\bibfield
   {journal} {\bibinfo  {journal} {Journal of Physics A: Mathematical, Nuclear
  and General}\ }\textbf {\bibinfo {volume} {7}},\ \bibinfo {pages} {1061}
  (\bibinfo {year} {1974})}\BibitemShut {NoStop}%
\bibitem [{\citenamefont {Teyssandier}\ and\ \citenamefont
  {Tourrenc}(1983)}]{Teyssandier&Tourrenc_Auxiliary_Scalars}%
  \BibitemOpen
  \bibfield  {author} {\bibinfo {author} {\bibfnamefont {P.}~\bibnamefont
  {Teyssandier}}\ and\ \bibinfo {author} {\bibfnamefont {P.}~\bibnamefont
  {Tourrenc}},\ }\href {\doibase 10.1063/1.525659} {\bibfield  {journal}
  {\bibinfo  {journal} {J.Math.Phys.}\ }\textbf {\bibinfo {volume} {24}},\
  \bibinfo {pages} {2793} (\bibinfo {year} {1983})}\BibitemShut {NoStop}%
\bibitem [{\citenamefont {Whitt}(1984)}]{Whitt_Auxiliary_Scalars}%
  \BibitemOpen
  \bibfield  {author} {\bibinfo {author} {\bibfnamefont {B.}~\bibnamefont
  {Whitt}},\ }\href {\doibase 10.1016/0370-2693(84)90332-0} {\bibfield
  {journal} {\bibinfo  {journal} {Physics Letters B}\ }\textbf {\bibinfo
  {volume} {145}},\ \bibinfo {pages} {176 } (\bibinfo {year}
  {1984})}\BibitemShut {NoStop}%
\bibitem [{\citenamefont {Benedetti}\ and\ \citenamefont
  {Guarnieri}(2013)}]{Benedetti:2013nya}%
  \BibitemOpen
  \bibfield  {author} {\bibinfo {author} {\bibfnamefont {D.}~\bibnamefont
  {Benedetti}}\ and\ \bibinfo {author} {\bibfnamefont {F.}~\bibnamefont
  {Guarnieri}},\ }\href@noop {} {\  (\bibinfo {year} {2013})},\ \Eprint
  {http://arxiv.org/abs/1311.1081} {arXiv:1311.1081 [hep-th]} \BibitemShut
  {NoStop}%
\bibitem [{\citenamefont {Sotiriou}\ and\ \citenamefont
  {Faraoni}(2010)}]{Sotiriou:2008rp}%
  \BibitemOpen
  \bibfield  {author} {\bibinfo {author} {\bibfnamefont {T.~P.}\ \bibnamefont
  {Sotiriou}}\ and\ \bibinfo {author} {\bibfnamefont {V.}~\bibnamefont
  {Faraoni}},\ }\href {\doibase 10.1103/RevModPhys.82.451} {\bibfield
  {journal} {\bibinfo  {journal} {Rev.Mod.Phys.}\ }\textbf {\bibinfo {volume}
  {82}},\ \bibinfo {pages} {451} (\bibinfo {year} {2010})},\ \Eprint
  {http://arxiv.org/abs/0805.1726} {arXiv:0805.1726 [gr-qc]} \BibitemShut
  {NoStop}%
\bibitem [{\citenamefont {De~Felice}\ and\ \citenamefont
  {Tsujikawa}(2010)}]{DeFelice:2010aj}%
  \BibitemOpen
  \bibfield  {author} {\bibinfo {author} {\bibfnamefont {A.}~\bibnamefont
  {De~Felice}}\ and\ \bibinfo {author} {\bibfnamefont {S.}~\bibnamefont
  {Tsujikawa}},\ }\href@noop {} {\bibfield  {journal} {\bibinfo  {journal}
  {Living Rev.Rel.}\ }\textbf {\bibinfo {volume} {13}},\ \bibinfo {pages} {3}
  (\bibinfo {year} {2010})},\ \Eprint {http://arxiv.org/abs/1002.4928}
  {arXiv:1002.4928 [gr-qc]} \BibitemShut {NoStop}%
\bibitem [{\citenamefont {Guberina}\ \emph {et~al.}(2003)\citenamefont
  {Guberina}, \citenamefont {Horvat},\ and\ \citenamefont
  {Stefancic}}]{Guberina:2002wt}%
  \BibitemOpen
  \bibfield  {author} {\bibinfo {author} {\bibfnamefont {B.}~\bibnamefont
  {Guberina}}, \bibinfo {author} {\bibfnamefont {R.}~\bibnamefont {Horvat}}, \
  and\ \bibinfo {author} {\bibfnamefont {H.}~\bibnamefont {Stefancic}},\ }\href
  {\doibase 10.1103/PhysRevD.67.083001} {\bibfield  {journal} {\bibinfo
  {journal} {Phys. Rev.}\ }\textbf {\bibinfo {volume} {D67}},\ \bibinfo {pages}
  {083001} (\bibinfo {year} {2003})},\ \Eprint
  {http://arxiv.org/abs/hep-ph/0211184} {arXiv:hep-ph/0211184} \BibitemShut
  {NoStop}%
\bibitem [{\citenamefont {Shapiro}\ \emph {et~al.}(2005)\citenamefont
  {Shapiro}, \citenamefont {Sola},\ and\ \citenamefont
  {Stefancic}}]{Shapiro:2004ch}%
  \BibitemOpen
  \bibfield  {author} {\bibinfo {author} {\bibfnamefont {I.~L.}\ \bibnamefont
  {Shapiro}}, \bibinfo {author} {\bibfnamefont {J.}~\bibnamefont {Sola}}, \
  and\ \bibinfo {author} {\bibfnamefont {H.}~\bibnamefont {Stefancic}},\ }\href
  {\doibase 10.1088/1475-7516/2005/01/012} {\bibfield  {journal} {\bibinfo
  {journal} {JCAP}\ }\textbf {\bibinfo {volume} {0501}},\ \bibinfo {pages}
  {012} (\bibinfo {year} {2005})},\ \Eprint
  {http://arxiv.org/abs/hep-ph/0410095} {arXiv:hep-ph/0410095} \BibitemShut
  {NoStop}%
\bibitem [{\citenamefont {Babic}\ \emph {et~al.}(2005)\citenamefont {Babic},
  \citenamefont {Guberina}, \citenamefont {Horvat},\ and\ \citenamefont
  {Stefancic}}]{Babic:2004ev}%
  \BibitemOpen
  \bibfield  {author} {\bibinfo {author} {\bibfnamefont {A.}~\bibnamefont
  {Babic}}, \bibinfo {author} {\bibfnamefont {B.}~\bibnamefont {Guberina}},
  \bibinfo {author} {\bibfnamefont {R.}~\bibnamefont {Horvat}}, \ and\ \bibinfo
  {author} {\bibfnamefont {H.}~\bibnamefont {Stefancic}},\ }\href {\doibase
  10.1103/PhysRevD.71.124041} {\bibfield  {journal} {\bibinfo  {journal} {Phys.
  Rev.}\ }\textbf {\bibinfo {volume} {D71}},\ \bibinfo {pages} {124041}
  (\bibinfo {year} {2005})},\ \Eprint {http://arxiv.org/abs/astro-ph/0407572}
  {arXiv:astro-ph/0407572} \BibitemShut {NoStop}%
\bibitem [{\citenamefont {Domazet}\ and\ \citenamefont
  {Stefancic}(2010)}]{Domazet:2010bk}%
  \BibitemOpen
  \bibfield  {author} {\bibinfo {author} {\bibfnamefont {S.}~\bibnamefont
  {Domazet}}\ and\ \bibinfo {author} {\bibfnamefont {H.}~\bibnamefont
  {Stefancic}},\ }\href@noop {} {\  (\bibinfo {year} {2010})},\ \Eprint
  {http://arxiv.org/abs/1010.3585} {arXiv:1010.3585 [gr-qc]} \BibitemShut
  {NoStop}%
\bibitem [{\citenamefont {Bonanno}\ and\ \citenamefont
  {Reuter}(2002{\natexlab{a}})}]{Bonanno:2001xi}%
  \BibitemOpen
  \bibfield  {author} {\bibinfo {author} {\bibfnamefont {A.}~\bibnamefont
  {Bonanno}}\ and\ \bibinfo {author} {\bibfnamefont {M.}~\bibnamefont
  {Reuter}},\ }\href {\doibase 10.1103/PhysRevD.65.043508} {\bibfield
  {journal} {\bibinfo  {journal} {Phys. Rev.}\ }\textbf {\bibinfo {volume}
  {D65}},\ \bibinfo {pages} {043508} (\bibinfo {year} {2002}{\natexlab{a}})},\
  \Eprint {http://arxiv.org/abs/hep-th/0106133} {arXiv:hep-th/0106133}
  \BibitemShut {NoStop}%
\bibitem [{\citenamefont {Bonanno}\ and\ \citenamefont
  {Reuter}(2002{\natexlab{b}})}]{Bonanno:2001hi}%
  \BibitemOpen
  \bibfield  {author} {\bibinfo {author} {\bibfnamefont {A.}~\bibnamefont
  {Bonanno}}\ and\ \bibinfo {author} {\bibfnamefont {M.}~\bibnamefont
  {Reuter}},\ }\href {\doibase 10.1016/S0370-2693(01)01522-2} {\bibfield
  {journal} {\bibinfo  {journal} {Phys. Lett.}\ }\textbf {\bibinfo {volume}
  {B527}},\ \bibinfo {pages} {9} (\bibinfo {year} {2002}{\natexlab{b}})},\
  \Eprint {http://arxiv.org/abs/astro-ph/0106468} {arXiv:astro-ph/0106468}
  \BibitemShut {NoStop}%
\bibitem [{\citenamefont {Bonanno}(2011)}]{Bonanno:2009nj}%
  \BibitemOpen
  \bibfield  {author} {\bibinfo {author} {\bibfnamefont {A.}~\bibnamefont
  {Bonanno}},\ }\href@noop {} {\bibfield  {journal} {\bibinfo  {journal} {PoS}\
  }\textbf {\bibinfo {volume} {CLAQG08}},\ \bibinfo {pages} {008} (\bibinfo
  {year} {2011})},\ \Eprint {http://arxiv.org/abs/0911.2727} {arXiv:0911.2727
  [hep-th]} \BibitemShut {NoStop}%
\bibitem [{\citenamefont {Bonanno}\ \emph {et~al.}(2011)\citenamefont
  {Bonanno}, \citenamefont {Contillo},\ and\ \citenamefont
  {Percacci}}]{Bonanno:2010bt}%
  \BibitemOpen
  \bibfield  {author} {\bibinfo {author} {\bibfnamefont {A.}~\bibnamefont
  {Bonanno}}, \bibinfo {author} {\bibfnamefont {A.}~\bibnamefont {Contillo}}, \
  and\ \bibinfo {author} {\bibfnamefont {R.}~\bibnamefont {Percacci}},\ }\href
  {\doibase 10.1088/0264-9381/28/14/145026} {\bibfield  {journal} {\bibinfo
  {journal} {Class.Quant.Grav.}\ }\textbf {\bibinfo {volume} {28}},\ \bibinfo
  {pages} {145026} (\bibinfo {year} {2011})},\ \Eprint
  {http://arxiv.org/abs/1006.0192} {arXiv:1006.0192 [gr-qc]} \BibitemShut
  {NoStop}%
\bibitem [{\citenamefont {Bonanno}(2012)}]{Bonanno:2012jy}%
  \BibitemOpen
  \bibfield  {author} {\bibinfo {author} {\bibfnamefont {A.}~\bibnamefont
  {Bonanno}},\ }\href@noop {} {\  (\bibinfo {year} {2012})},\ \bibinfo {note}
  {5 pages, to appear as a Rapid Communication in Physical Review D},\ \Eprint
  {http://arxiv.org/abs/1203.1962} {arXiv:1203.1962 [hep-th]} \BibitemShut
  {NoStop}%
\bibitem [{\citenamefont {Frolov}\ and\ \citenamefont
  {Guo}(2011)}]{Frolov:2011ys}%
  \BibitemOpen
  \bibfield  {author} {\bibinfo {author} {\bibfnamefont {A.~V.}\ \bibnamefont
  {Frolov}}\ and\ \bibinfo {author} {\bibfnamefont {J.-Q.}\ \bibnamefont
  {Guo}},\ }\href@noop {} {\  (\bibinfo {year} {2011})},\ \Eprint
  {http://arxiv.org/abs/1101.4995} {arXiv:1101.4995 [astro-ph.CO]} \BibitemShut
  {NoStop}%
\bibitem [{\citenamefont {Hindmarsh}\ \emph {et~al.}(2011)\citenamefont
  {Hindmarsh}, \citenamefont {Litim},\ and\ \citenamefont
  {Rahmede}}]{Hind-Litim_Rahme}%
  \BibitemOpen
  \bibfield  {author} {\bibinfo {author} {\bibfnamefont {M.}~\bibnamefont
  {Hindmarsh}}, \bibinfo {author} {\bibfnamefont {D.}~\bibnamefont {Litim}}, \
  and\ \bibinfo {author} {\bibfnamefont {C.}~\bibnamefont {Rahmede}},\ }\href
  {\doibase 10.1088/1475-7516/2011/07/019} {\bibfield  {journal} {\bibinfo
  {journal} {JCAP}\ }\textbf {\bibinfo {volume} {1107}},\ \bibinfo {pages}
  {019} (\bibinfo {year} {2011})},\ \Eprint {http://arxiv.org/abs/1101.5401}
  {arXiv:1101.5401 [gr-qc]} \BibitemShut {NoStop}%
\bibitem [{\citenamefont {Koch}\ and\ \citenamefont
  {Ramirez}(2011)}]{Koch:2010nn}%
  \BibitemOpen
  \bibfield  {author} {\bibinfo {author} {\bibfnamefont {B.}~\bibnamefont
  {Koch}}\ and\ \bibinfo {author} {\bibfnamefont {I.}~\bibnamefont {Ramirez}},\
  }\href {\doibase 10.1088/0264-9381/28/5/055008} {\bibfield  {journal}
  {\bibinfo  {journal} {Class.Quant.Grav.}\ }\textbf {\bibinfo {volume} {28}},\
  \bibinfo {pages} {055008} (\bibinfo {year} {2011})},\ \Eprint
  {http://arxiv.org/abs/1010.2799} {arXiv:1010.2799 [gr-qc]} \BibitemShut
  {NoStop}%
\bibitem [{\citenamefont {Koch}\ \emph {et~al.}(2015)\citenamefont {Koch},
  \citenamefont {Rioseco},\ and\ \citenamefont {Contreras}}]{Koch:2014joa}%
  \BibitemOpen
  \bibfield  {author} {\bibinfo {author} {\bibfnamefont {B.}~\bibnamefont
  {Koch}}, \bibinfo {author} {\bibfnamefont {P.}~\bibnamefont {Rioseco}}, \
  and\ \bibinfo {author} {\bibfnamefont {C.}~\bibnamefont {Contreras}},\ }\href
  {\doibase 10.1103/PhysRevD.91.025009} {\bibfield  {journal} {\bibinfo
  {journal} {Phys.Rev.}\ }\textbf {\bibinfo {volume} {D91}},\ \bibinfo {pages}
  {025009} (\bibinfo {year} {2015})},\ \Eprint {http://arxiv.org/abs/1409.4443}
  {arXiv:1409.4443 [hep-th]} \BibitemShut {NoStop}%
\bibitem [{\citenamefont {Coleman}\ and\ \citenamefont
  {Weinberg}(1973)}]{Coleman:1973jx}%
  \BibitemOpen
  \bibfield  {author} {\bibinfo {author} {\bibfnamefont {S.~R.}\ \bibnamefont
  {Coleman}}\ and\ \bibinfo {author} {\bibfnamefont {E.~J.}\ \bibnamefont
  {Weinberg}},\ }\href {\doibase 10.1103/PhysRevD.7.1888} {\bibfield  {journal}
  {\bibinfo  {journal} {Phys.Rev.}\ }\textbf {\bibinfo {volume} {D7}},\
  \bibinfo {pages} {1888} (\bibinfo {year} {1973})}\BibitemShut {NoStop}%
\bibitem [{\citenamefont {Mukhanov}\ \emph {et~al.}(1992)\citenamefont
  {Mukhanov}, \citenamefont {Feldman},\ and\ \citenamefont
  {Brandenberger}}]{Mukhanov:1990me}%
  \BibitemOpen
  \bibfield  {author} {\bibinfo {author} {\bibfnamefont {V.~F.}\ \bibnamefont
  {Mukhanov}}, \bibinfo {author} {\bibfnamefont {H.}~\bibnamefont {Feldman}}, \
  and\ \bibinfo {author} {\bibfnamefont {R.~H.}\ \bibnamefont
  {Brandenberger}},\ }\href {\doibase 10.1016/0370-1573(92)90044-Z} {\bibfield
  {journal} {\bibinfo  {journal} {Phys.Rept.}\ }\textbf {\bibinfo {volume}
  {215}},\ \bibinfo {pages} {203} (\bibinfo {year} {1992})}\BibitemShut
  {NoStop}%
\bibitem [{\citenamefont {{Amendola}}\ \emph {et~al.}(1992)\citenamefont
  {{Amendola}}, \citenamefont {{Capozziello}}, \citenamefont {{Litterio}},\
  and\ \citenamefont {{Occhionero}}}]{Amend1992PRD45417A}%
  \BibitemOpen
  \bibfield  {author} {\bibinfo {author} {\bibfnamefont {L.}~\bibnamefont
  {{Amendola}}}, \bibinfo {author} {\bibfnamefont {S.}~\bibnamefont
  {{Capozziello}}}, \bibinfo {author} {\bibfnamefont {M.}~\bibnamefont
  {{Litterio}}}, \ and\ \bibinfo {author} {\bibfnamefont {F.}~\bibnamefont
  {{Occhionero}}},\ }\href {\doibase 10.1103/PhysRevD.45.417} {\bibfield
  {journal} {\bibinfo  {journal} {\prd}\ }\textbf {\bibinfo {volume} {45}},\
  \bibinfo {pages} {417} (\bibinfo {year} {1992})}\BibitemShut {NoStop}%
\bibitem [{\citenamefont {{Maeda}}(1988)}]{Maeda1988PRD37858M}%
  \BibitemOpen
  \bibfield  {author} {\bibinfo {author} {\bibfnamefont {K.-I.}\ \bibnamefont
  {{Maeda}}},\ }\href {\doibase 10.1103/PhysRevD.37.858} {\bibfield  {journal}
  {\bibinfo  {journal} {\prd}\ }\textbf {\bibinfo {volume} {37}},\ \bibinfo
  {pages} {858} (\bibinfo {year} {1988})}\BibitemShut {NoStop}%
\bibitem [{\citenamefont {Hwang}\ and\ \citenamefont
  {Noh}(2005)}]{PhysRevD.71.063536}%
  \BibitemOpen
  \bibfield  {author} {\bibinfo {author} {\bibfnamefont {J.-c.}\ \bibnamefont
  {Hwang}}\ and\ \bibinfo {author} {\bibfnamefont {H.}~\bibnamefont {Noh}},\
  }\href {\doibase 10.1103/PhysRevD.71.063536} {\bibfield  {journal} {\bibinfo
  {journal} {Phys. Rev. D}\ }\textbf {\bibinfo {volume} {71}},\ \bibinfo
  {pages} {063536} (\bibinfo {year} {2005})}\BibitemShut {NoStop}%
\bibitem [{\citenamefont {Hwang}\ and\ \citenamefont
  {Noh}(1996)}]{PhysRevD.54.1460}%
  \BibitemOpen
  \bibfield  {author} {\bibinfo {author} {\bibfnamefont {J.-c.}\ \bibnamefont
  {Hwang}}\ and\ \bibinfo {author} {\bibfnamefont {H.}~\bibnamefont {Noh}},\
  }\href {\doibase 10.1103/PhysRevD.54.1460} {\bibfield  {journal} {\bibinfo
  {journal} {Phys. Rev. D}\ }\textbf {\bibinfo {volume} {54}},\ \bibinfo
  {pages} {1460} (\bibinfo {year} {1996})}\BibitemShut {NoStop}%
\bibitem [{\citenamefont {Litim}(2001)}]{Litim:2001up}%
  \BibitemOpen
  \bibfield  {author} {\bibinfo {author} {\bibfnamefont {D.~F.}\ \bibnamefont
  {Litim}},\ }\href {\doibase 10.1103/PhysRevD.64.105007} {\bibfield  {journal}
  {\bibinfo  {journal} {Phys.Rev.}\ }\textbf {\bibinfo {volume} {D64}},\
  \bibinfo {pages} {105007} (\bibinfo {year} {2001})},\ \Eprint
  {http://arxiv.org/abs/hep-th/0103195} {arXiv:hep-th/0103195 [hep-th]}
  \BibitemShut {NoStop}%
\bibitem [{\citenamefont {Litim}(2000{\natexlab{b}})}]{Litim:2000ci}%
  \BibitemOpen
  \bibfield  {author} {\bibinfo {author} {\bibfnamefont {D.~F.}\ \bibnamefont
  {Litim}},\ }\href {\doibase 10.1016/S0370-2693(00)00748-6} {\bibfield
  {journal} {\bibinfo  {journal} {Phys.Lett.}\ }\textbf {\bibinfo {volume}
  {B486}},\ \bibinfo {pages} {92} (\bibinfo {year} {2000}{\natexlab{b}})},\
  \Eprint {http://arxiv.org/abs/hep-th/0005245} {arXiv:hep-th/0005245 [hep-th]}
  \BibitemShut {NoStop}%
\bibitem [{\citenamefont {York}(1973)}]{York:1973ia}%
  \BibitemOpen
  \bibfield  {author} {\bibinfo {author} {\bibfnamefont {J.}~\bibnamefont
  {York}, \bibfnamefont {James~W.}},\ }\href {\doibase 10.1063/1.1666338}
  {\bibfield  {journal} {\bibinfo  {journal} {J.Math.Phys.}\ }\textbf {\bibinfo
  {volume} {14}},\ \bibinfo {pages} {456} (\bibinfo {year} {1973})}\BibitemShut
  {NoStop}%
\bibitem [{\citenamefont {chan Hwang}\ and\ \citenamefont
  {Noh}(2001)}]{Hwang200113}%
  \BibitemOpen
  \bibfield  {author} {\bibinfo {author} {\bibfnamefont {J.}~\bibnamefont {chan
  Hwang}}\ and\ \bibinfo {author} {\bibfnamefont {H.}~\bibnamefont {Noh}},\
  }\href {\doibase http://dx.doi.org/10.1016/S0370-2693(01)00404-X} {\bibfield
  {journal} {\bibinfo  {journal} {Physics Letters B}\ }\textbf {\bibinfo
  {volume} {506}},\ \bibinfo {pages} {13 } (\bibinfo {year}
  {2001})}\BibitemShut {NoStop}%
\end{thebibliography}%

\end{document}